\documentclass{article}
\usepackage{relsize}
\usepackage{graphicx}
\usepackage{tikz}
\usepackage{tikz-cd}
\usetikzlibrary{arrows.meta,decorations.pathreplacing,calc,positioning,fit}
\tikzset{>={Latex[length=2mm]},
  lab/.style={font=\small},
  tiny/.style={font=\footnotesize},
  box/.style={draw,rounded corners,thick,inner sep=3pt,fill=black!2},
  base/.style={draw,rounded corners,thick,inner sep=3pt,fill=black!1}}
\usepackage{authblk}
\usepackage{amsmath}
 \usepackage{natbib}
\usepackage{pgfplots}
\usepackage{multirow}
\usepackage{array}
\pgfplotsset{compat=1.18}
\usepackage{caption}

\usepackage{subcaption}
\usepackage{adjustbox}
\usepackage{tabularx}
\usepackage[hidelinks]{hyperref}
\usepackage[english]{babel}

\usepackage{xcolor}
\usepackage{float}
\usepackage{newunicodechar}
\newunicodechar{−}{\ensuremath{-}}
\usepackage{booktabs}
\usepackage{slashed,epsfig,amsmath,amssymb,enumitem} \usepackage[papersize={8.5in,11in}]{geometry}
   \usepackage{bm}
\usepackage{graphicx}

\newcommand{\tr}{\operatorname{tr}}
\usepackage[hidelinks]{hyperref}
\newcommand{\be}{\begin{equation}}
\newcommand{\ee}{\end{equation}}

\usepackage[utf8]{inputenc}
\usepackage{epigraph}
\begin{document}

\title{On the Complexified Spacetime Manifold Mapping of AdS to dS}

\author[1 2]{J. W. Moffat}

\author[1 3]{E. J. Thompson}

\affil[1]{Perimeter Institute for Theoretical Physics, Waterloo, Ontario N2L 2Y5, Canada}

\affil[2]{Department of Physics and Astronomy, University of Waterloo, Waterloo,
Ontario N2L 3G1, Canada}

\affil[3]{Department of Physics and Astronomy, Trent University, Peterborough, 
Ontario K9L 0G2, Canada}

\maketitle

\begin{abstract}
In a complex manifold, one can bridge anti-de Sitter and de Sitter spacetimes via analytic continuation, preserving geometric invariants and regularity, avoiding singularities during the AdS–dS transition. It unifies gravitational and gauge interactions under a complexified symmetry group, maintaining bulk unitarity for both AdS and dS. Boundary unitarity is upheld in AdS but not in dS due to the spacelike conformal boundary. The theory uses holographic principles like the MacDowell–Mansouri and Quantum Extremal Surface prescriptions to align entanglement and black hole entropy with AdS/CFT and general relativity. HUFT provides insights into AdS and dS holography, the cosmological constant, and quantum gravity unitarity and entanglement.
\end{abstract}

\section{Introduction}

A central tension in quantum gravity is the asymptotic and causal dichotomy between anti–de Sitter (AdS) and de Sitter (dS) spacetimes. AdS admits a timelike conformal boundary  and a unitary boundary description \cite{Maldacena1998,Witten1998,GKP1998,Aharony2000}, while dS possesses spacelike future and past screens, lacks global in and out states, and generically defies a unitary boundary Hilbert space \cite{Strominger2001,Anninos2012}. In this work we show that both geometries arise naturally and analytically within a single holomorphic framework built on a complex four–dimensional ambient manifold \(M_{\mathbb C}\) with coordinates \(z^\mu=x^\mu+i\,y^\mu\) \cite{MT:HUFT-EPJC,MT:Invariant}. Physical observables live on the real slice \(y^\mu=0\), while the auxiliary directions organize a Picard–Lefschetz contour for the path integral that selects the real slice as the dominant saddle \cite{MT:HUFT-EPJC, Witten2010QM,Witten2011CS}.

The AdS–dS bridge follows from analytic continuation in a holomorphic cosmological constant, or equivalently, a complex rotation of the curvature scale \cite{SleightTaronna2019, Skenderis2010, McFaddenSkenderis2010}. We give explicit coordinate continuations and prove that all scalar curvature invariants remain finite along the continuation, maximally symmetric vacua satisfy \(R_{\mu\nu}=\frac{2\Lambda}{d-1}\,g_{\mu\nu}\), so invariants are polynomial in \(\Lambda\) and even under \(\Lambda\to-\Lambda\). The Gibbons–Hawking–York term flips sign through the change in causal character of the cutoff surface, but the variational problem and Brown–York machinery remain well-posed \cite{York1972,GibbonsHawking1977,BrownYork1993,BalasubramanianKraus1999}. Black-hole saddles depend smoothly on \(\Lambda\), horizons move continuously, and curvature scalars stay finite, so no new singularities are introduced by the AdS\(\leftrightarrow\)dS rotation \cite{Nariai1951}.

We formulate quantization on the real slice with a Schwinger–Keldysh (CPT) contour \cite{Schwinger1961,Keldysh1964}. Bulk unitarity then follows from the largest-time identity \(Z[J,J]=1\), yielding standard cutting rules and the optical theorem \cite{Cutkosky1960}. Entire–function dressing leaves cuts and phase space unaltered\cite{Moffat:FiniteNonlocal1990,Tomboulis1997,Biswas2012,Talaganis2015,Buoninfante2018,MT:FiniteHolomorphicQFT}. By contrast, boundary unitarity holds only on the AdS side, Euclidean AdS on-shell functionals are reflection positive and reconstruct a unitary Lorentzian CFT \cite{OsterwalderSchrader1973,Skenderis2002}, whereas the dS late-time functional \(W_{\mathrm dS}\) acquires a universal phase that generically violates reflection positivity on spacelike screens \cite{MaldacenaPimentel2011,BunchDavies1978,Allen1985,Mottola1985}. Nevertheless, physically measurable \textbf{in–in} observables in dS remain unitary, and static-patch correlators satisfy KMS thermal periodicity\footnote{Ryogo Kubo introduced the condition in 1957, Paul C. Martin and Julian Schwinger used it in 1959 to define thermodynamic Green's functions, [2] and Rudolf Haag, Marinus Winnink and Nico Hugenholtz used the condition in 1967 to define equilibrium states and called it the KMS condition.} at \(T=H/2\pi\) \cite{GibbonsHawking1977}. In de Sitter cosmology, late-time correlators also act as a cosmological collider, where massive fields imprint oscillatory non-Gaussian signals \cite{ArkaniHamedMaldacena2015}

Finally, we develop holography in this holomorphic setting. On the AdS slice we recover standard thermodynamics and entanglement through Ryu–Takayanagi (RT) for static spacetime, Hubeny–Rangamani–Takayanagi (HRT) for covariant or general spacetimes, Quantum Extremal Surface (QES) and the generalized-entropy formula \cite{RyuTakayanagi2006,HRT2007,LewkowyczMaldacena2013,EngelhardtWall2015}. Through the analytic bridge, these structures carry over to dS as state-preparation holography, generalized entropies of horizons and static-patch thermality follow, while the lack of a unitary boundary CFT is traced to the spacelike nature of \(\mathcal I^{\pm}\) \cite{Strominger2001,Anninos2012, Skenderis2010, McFaddenSkenderis2010}.

HUFT shows that AdS and dS are not two unrelated universes but that they are two parts of the same geometric object. In a complexified spacetime, you can rotate the single dial that sets curvature, the cosmological constant and smoothly move from AdS a region of negative curvature, timelike boundary to dS a region with positive curvature, spacelike screens. Because the rotation is analytic, curvature scalars stay finite and there are no new singularities that appear. Gravity on the real slice is just MacDowell–Mansouri \cite{MacDowellMansouri1977}, a gauge-theory form of GR with $\Lambda$, and the Standard Model lives there too, with ordinary conservation laws so no charge or energy leaking into the auxiliary directions. Bulk quantum evolution is unitary on either slice so what changes is the kind of holography one does, a unitary boundary CFT in AdS versus a state-preparation wavefunctional in dS.

This turns the AdS–dS either–or into a single, controlled both. AdS and dS become two real slices of one holomorphic theory, with an analytic bridge that introduces no new singularities. That bridge lets us port AdS tools such as entanglement or Quantum Extremal Surface, thermodynamics, black-hole methods, directly into our $\Lambda>0$ universe via state-preparation, giving a practical calculational pipeline for cosmology. It also clarifies unitarity where bulk unitarity holds on both sides; while dS’s boundary non-unitarity is a causal feature of spacelike screens, not a flaw. And because GR and the Standard Model live consistently on the same 3+1 slice and the flat limit respects the Coleman–Mandula theorem~\cite{ColemanMandula1967}, the framework stays tied to real-world physics and testable predictions.

HUFT provides a unified, calculable framework in which AdS and dS are two regular real slices of a single complex ambient theory, reconciling bulk unitarity, black-hole thermodynamics, and holographic entanglement across the AdS–dS divide.

\section{The Holomorphic Unified Field Theory}
\label{sec:huft-overview}
HUFT aims to put all of known micro- and macro-physics into one geometric package and make it mathematically clean, we wanted it to be a single, symmetry-exact, UV-finite geometric theory whose real-slice limit reproduces General Relativity and the full Standard Model, including chiral fermions and their masses \cite{MT:HUFT-EPJC,MT:Invariant,Moffat:FiniteNonlocal1990,MT:FiniteHolomorphicQFT, Moffat1, Moffat2, MT:SMmass,MT:SL2C,MT:ReplyToCline, Moffat:UVcompleteQG2011,Moffat:QGCCP2014, Cline:CommentHUFT}. HUFT is formulated on a complexified four–manifold \(M_{\mathbb C}\) with local coordinates:
\begin{equation}
  z^\mu  =  x^\mu + i\,y^\mu, \qquad \mu=0,1,2,3,
\end{equation}
equipped with a Hermitian metric:
\begin{equation}
  g_{\mu\nu}(z,\bar z) = h_{\mu\nu}(x,y)  +  i\,B_{\mu\nu}(x,y),
  \qquad h_{\mu\nu}=h_{\nu\mu},  B_{\mu\nu}=-B_{\nu\mu}.
\end{equation}
Physical observables live on the real slice where \(y^\mu=0\), where the symmetric part \(h_{\mu\nu}\) is the Lorentzian spacetime metric and the antisymmetric part \(B_{\mu\nu}\) is identified with the gauge-sector curvature packaged as a spacetime two–tensor.

The unified geometric data on the real slice comes when we let \(M\) denote the real slice, so the kinematics are encoded by the product principal bundle:
\begin{equation}
  P_{\text{tot}}  =  P_{\mathrm{Spin}(1,3)} \times_{M} P_G,
  \qquad H  =  \mathrm{Spin}(1,3)\times G,
\end{equation}
with a single \(H\)-connection:
\begin{equation}
  \mathcal{A}  =  (\omega, A),
  \qquad \omega\in\Omega^1(M,\mathfrak{so}(1,3)),  A\in\Omega^1(M,\mathfrak{g}),
\end{equation}
and curvature given by:
\begin{equation}
  \mathcal{F}  =  d\mathcal{A}+\mathcal{A}\wedge\mathcal{A}
   =  (R,\,F), \quad
  R\in\Omega^2(M,\mathfrak{so}(1,3)),  F\in\Omega^2(M,\mathfrak{g}).
\end{equation}
We on the real slice identify the antisymmetric part of the Hermitian metric with the Yang–Mills curvature:
\begin{equation}
  B_{\mu\nu}  \equiv  \alpha\,\,F^a_{\mu\nu},
  \label{eq:B-equals-F}
\end{equation}
where $\kappa_{ab}$ is the Killing form on $\mathfrak{g}$ and $\alpha$ is a dimensionful constant, so that the Hermitian packaging:
\begin{equation}
  g_{\mu\nu}\big|_{y=0}  =  h_{\mu\nu}  +  i\,\alpha\,F^a_{\mu\nu},
  \label{eq:Hermitian-packaging}
\end{equation}
encodes both spacetime geometry and gauge curvature in a single tensorial object. After choosing the Picard–Lefschetz contour, a minimal Diff\((M)\)\(\times\)G–invariant action on the real slice is:
\begin{equation}
  S[h,A,\Psi]
   = 
  \int_M\! d^4x\,\sqrt{|h|}
  \left[
    \frac{1}{2\kappa}\,R(h)
     - \frac{1}{4}\,\langle F,F\rangle_h
     +  \mathcal{L}_{\text{matter}}(\Psi;h,A)
  \right],
  \label{eq:unified-action}
\end{equation}
where \(\langle F,F\rangle_h := \kappa_{ab}\,F^a_{\mu\nu}F^{b\,\mu\nu}\) with indices raised by \(h^{\mu\nu}\), variations yield:
\begin{align}
  G_{\mu\nu}(h) &= \kappa\left(T^{\text{YM}}_{\mu\nu}+T^{\text{matter}}_{\mu\nu}\right),
  \label{eq:Einstein-eq} \\
  D_A({}^{\!*}_h F) &= J_{\text{matter}},
  \label{eq:YM-eq}
\end{align}
with Bianchi identities and conservation laws:
\begin{equation}
  D_\omega R=0,\qquad D_A F=0,\qquad \nabla_\mu T^{\mu}{}_{\nu}=0,\qquad D_A J_{\text{matter}}=0.
  \label{eq:Bianchi-conservation}
\end{equation}
Equations \eqref{eq:Einstein-eq}–\eqref{eq:Bianchi-conservation} show that a single symmetry principle, \text{Diff}\(\times\)G produces gravity, Yang–Mills, and their conserved currents, the Hermitian map \eqref{eq:Hermitian-packaging} is an on–shell equivalence between the holomorphic and real-slice descriptions.

The holomorphic path integral over \(g\) and \(\mathcal{A}\) is defined on suitable middle–dimensional cycles in field space. Choosing the Picard–Lefschetz contour that passes through the real saddle \(y^\mu=0\) gives:
\begin{equation}
  \int_{\mathcal{C}}\!\mathcal{D}g\,\mathcal{D}\mathcal{A}\,e^{i S_{\mathbb C}[g,\mathcal{A}]}
   \simeq 
  \int \!\mathcal{D}h\,\mathcal{D}A\,\mathcal{D}\Psi\,e^{i S[h,A,\Psi]},
\end{equation}
with subdominant saddles exponentially suppressed. Thus, the holomorphic theory reduces to the standard real action \eqref{eq:unified-action} without introducing extra degrees of freedom on the physical slice.

The internal bundle structure and minimality toward \(SU(5)\) comes when we let \(E\to M\) be the holomorphic vector bundle associated to \(P_G\). Assuming \(c_1(E)=0\) and a nowhere-vanishing holomorphic volume form \(\Omega_E\in H^{n,0}(E)\), the structure group reduces \(U(n)\to SU(n)\).
Chirality, anomaly cancellation, and charge integrality constrain \(n\), with the minimal consistent choice yielding \(n=5\):
\begin{equation}
  SU(5)\ \longrightarrow\ SU(3)_c\times SU(2)_L\times U(1)_Y,
\end{equation}
on the real slice. The detailed breaking pattern and hypercharge normalization follow from Chern–Weil integrality and the embedding of \(\mathfrak{su}(3)\oplus\mathfrak{su}(2)\oplus\mathfrak{u}(1)\subset\mathfrak{su}(5)\).

At the quantum level, HUFT employs an entire-function form factor built from covariant operators:
\begin{equation}
  F\!\left(\frac{D^2}{M_*^2}\right) = \exp\!\left(-\frac{D^2}{M_*^2}\right),
  \qquad D_\mu=\partial_\mu + (\omega_\mu, A_\mu),
  \label{eq:entire-regulator}
\end{equation}
inserted between gauge–covariant quantities in loops.
Because \(F\) is an analytic function of the covariant Laplacian, it transforms by conjugation, preserving Ward or Slavnov–Taylor identities.\footnote{Equivalently, one may use a proper–time heat–kernel representation with path-ordered Wilson lines to make covariance manifest. The entire form factor introduces no new poles, so unitarity is preserved while UV modes are exponentially damped.}
For example, the regulated gauge propagator in momentum space has the schematic form:
\begin{equation}
  \tilde{\Delta}_{\mu\nu}(p)
   = 
  \frac{e^{-p^2/M_*^2}}{p^2+i0^+}\,
  \Pi_{\mu\nu}^{\text{(T)}}(p)  +  \xi\,\frac{e^{-p^2/M_*^2}}{p^2+i0^+}\,\Pi_{\mu\nu}^{\text{(L)}}(p),
\end{equation}
and the one–loop vacuum polarization remains transverse:
\begin{equation}
  \Pi_{\mu\nu}(p) = \big(p^2\eta_{\mu\nu}-p_\mu p_\nu\big)\,\Pi_R(p^2;M_*),
\end{equation}
so a photon mass term is not generated. The gauge couplings then obey modified one–loop RGEs of the form:
\begin{equation}
  \mu\,\frac{d g_i}{d\mu}
   = 
  \frac{b_i}{16\pi^2}\,g_i^3\,
  \exp\!\Bigl(-\frac{\mu^2}{M_*^2}\Bigr),
  \qquad (b_1,b_2,b_3)=\Bigl(\tfrac{41}{10},-\tfrac{19}{6},-7\Bigr),
  \label{eq:rg-huft}
\end{equation}
which preserve asymptotic freedom for \(SU(3)\), soften UV growth, and can sharpen gauge-coupling unification tests.

Now one must ask what is conceptually new, we have single Hermitian tensor \(g_{\mu\nu}=h_{\mu\nu}+i\,B_{\mu\nu}\) packages gravity and gauge curvature; with the identification \eqref{eq:B-equals-F}, the antisymmetric sector is not an extra field but the Yang–Mills 2–form. One Diff\(\times\)G–invariant action \eqref{eq:unified-action} yields Einstein plus Yang–Mills plus matter and their conservation laws \eqref{eq:Bianchi-conservation}; the holomorphic path integral selects the real slice without ad hoc constraints. A gauge–covariant entire regulator \eqref{eq:entire-regulator} renders loops finite while preserving symmetries and masslessness of gauge bosons, leading to predictive, UV–softened RG flow \eqref{eq:rg-huft}.  In applications we set \(\alpha\) in \eqref{eq:B-equals-F} by dimensional analysis and phenomenological normalization absorbing group factors into \(\kappa_{ab}\); indices are always raised or lowered with \(h_{\mu\nu}\) on the real slice, and traces over \(\mathfrak{g}\) use \(\kappa_{ab}\).
\begin{equation}
    G_{\mu\nu}=\kappa T_{\mu\nu}, \quad DA(\ast F)=\ast J, \quad \nabla_\mu T^\mu_\nu=0, \quad D_\mu J^\mu =0,
\end{equation}
with the Bianchi identities $D_\omega R=0,\quad DAF=0$ providing the Noether relations.

\section{Complex Manifold Mapping of AdS to dS}

When mapping from AdS to dS, the sign of the cosmological constant term is either negative or positive. A timelike boundary is used for particle physics, while a spacelike boundary is used for cosmology \cite{Maldacena1998,Witten1998,Aharony2000,Strominger2001,Anninos2012}. Particle physics needs a setup where we can study evolution in time, so it lives in timelike boundary worlds, either AdS or flat. Cosmology deals with entire spacelike slices of the universe, especially at large times living on spacelike boundaries dS \cite{Anninos2012}. In holography, AdS/CFT naturally models quantum field theories, and a hypothetical dS/CFT is formulated on a spacelike surface, matching the end-of-universe idea in cosmology \cite{Maldacena1998,Strominger2001}. On $M_\mathbb{C}$, the Einstein field equations are given by:
\begin{equation}
    G_{\mu\nu}(z)+\Lambda h_{\mu\nu}(z)=\kappa T_{\mu\nu}^{(m)}(z),
\end{equation}
and on the real slice produce the $x$ dependent field equations.

We have in HUFT a way we can naturally bridge AdS to dS. First, we have the same contour but different backgrounds through Picard-Lefschetz with quantization through the holomorphic path integral. The integration cycle $C$ is chosen via Picard Lefschetz in the homology class of the real slice so that the real-slice saddle dominates while other thimbles give subleading corrections \cite{MT:HUFT-EPJC, TanizakiKoike2014,FeldbruggeLehnersTurok2017PRD,FeldbruggeLehnersTurok2017PRL}. When moving between AdS and dS is just which classical background one expands about. Inside HUFT, we treat the cosmological constant as a holomorphic parameter:
\begin{equation}
    \Lambda(\theta)=|\Lambda|e^{i\theta},
\end{equation}
where $\theta$ is the phase angle is just the phase angle of the cosmological constant in the complex 
$\Lambda$-plane, it is the knob used inside the holomorphic path integral on $M_{\mathbb{C}}$ to move between constant curvature saddles. We view AdS and dS as different real slice endpoints of the analytic family of $\theta$.
The physical observables live on the endpoints of $\theta$ and is only manifestly real at the phases:
\begin{equation}
    \theta=\pi \quad(\Lambda<0, \text{AdS}), \qquad \theta=0 \quad(\Lambda>0, \text{dS}),
\end{equation}
while we treat the intermediate phases $\theta\in (0,\pi)$ to correspond to complex saddles in the holomorphic path integral that you may traverse when deforming the contours through Picard-Lefschetz, but they are not themselves physical real slice backgrounds.

We can also model the bridge between AdS and dS in terms of curvature scales. We start by writing the (A)dS scales in $d+1$ bulk dimensions as:
\begin{equation}
\Lambda_{\text{AdS}}=\frac{-d(d-1)}{2L^2}, \qquad \Lambda_{\text{dS}}=\frac{d(d-1)}{2}H^2,
\end{equation}
in this, $d+1$ is the bulk spacetime dimension, $L$ is the AdS curvature radius, AdS has constant negative curvature; its sectional curvature is $\frac{-1}{L^2}$, and $H$ is the constant Hubble parameter of de Sitter; dS has sectional curvature of $H^2$. For any maximally symmetric space in $D=d+1$ dimensions\footnote{Physically spacetime is 3 space + 1 time, the bulk is physical 4D spacetime. The bulk we compute in is still 3+1D physical spacetime; using D=d+1 is just standard holographic bookkeeping, here we set d=3 for a 4D bulk.}:
\begin{equation}
R_{\mu\nu}
= \frac{2\Lambda}{D-2}\, g_{\mu\nu}
= \frac{2\Lambda}{d-1}\, g_{\mu\nu}.
\end{equation}
On the other hand:
\begin{equation}
R_{\mu\nu} = d\,\sigma\, g_{\mu\nu}, 
\qquad 
\sigma =
\begin{cases}
-\dfrac{1}{L^{2}}, & \text{(AdS)},\\[4pt]
+H^{2}, & \text{(dS)},
\end{cases}
\end{equation}
where $\sigma$ is the constant sectional curvature of a maximally symmetric spacetime. Equating this gives: \begin{equation}
\Lambda = \frac{d(d-1)}{2}\,\sigma,
\end{equation}
such as:
\begin{equation}
\Lambda_{\mathrm{AdS}} = -\frac{d(d-1)}{2L^{2}},
\qquad
\Lambda_{\mathrm{dS}} = \frac{d(d-1)}{2}\,H^{2}.
\end{equation}
A useful limit and identification for these equations; for flat Minkowski limit: 
$L\to\infty$ (AdS) or $H\to 0 (\text{dS})$, $\Rightarrow \Lambda\to 0$, with: 
\be
H=\frac{1}{L_{\mathrm{dS}}}, \qquad \Lambda_{\mathrm{dS}}=\dfrac{d(d-1)}{2L_{\mathrm{dS}}^{2}}.
\ee
We bridge by analytic continuation given by $L \mapsto i/H$, equivalently $\Lambda \mapsto -\Lambda$, this flips the sign of the curvature between AdS and dS \cite{SleightTaronna2019,Skenderis2010, McFaddenSkenderis2010}. The Fefferman–Graham to flat-slicing map $z\rightarrow i\eta$, \footnote{Where $\eta$ is the de Sitter conformal time coordinate and $z$ is the Poincar\'e radial coordinate of AdS.} realizes this continuation explicitly \cite{Skenderis2010}.
A single complex rotation of the length scale inside $M_\mathbb{C}$:
\begin{equation}
    L\to e^{i\alpha}L,
\end{equation}
where $\alpha$ is the generally complex rotation parameter that tells you how far you rotate the AdS or dS length scale in the complex plane of the complexified spacetime manifold:
\begin{equation}
L(\alpha)=e^{i\alpha}L_0,\qquad \forall \space \space \alpha \space \in \space \mathbb{C}.
\end{equation}
If $\alpha$ is real and a pure phase:
\begin{equation}
\frac{1}{L^2}  \mapsto  \frac{1}{\big(e^{i\alpha}L_0\big)^2}
= e^{-2i\alpha}\,\frac{1}{L_0^2}.
\end{equation}
Choosing $\alpha=\tfrac{\pi}{2}$ gives $L\to iL$ and flips the sign of the curvature:
\begin{equation}
\sigma_{\mathrm{AdS}}=-\frac{1}{L_0^{2}}
 \longrightarrow 
+\frac{1}{L_0^{2}}=\sigma_{\mathrm{dS}},
\qquad \text{equivalently}\quad \Lambda \to -\Lambda.
\end{equation}
If $\alpha=a+ib$ is complex:
\begin{equation}
L\to e^{ia-b}L_0
\quad\Rightarrow\quad
\left|\frac{1}{L^{2}}\right| = e^{2b}\,\frac{1}{L_0^{2}},
\end{equation}
and the cosmological constant transforms as
\begin{equation}
\Lambda(\alpha)=e^{-2i\alpha}\,\Lambda_0 .
\end{equation}
To keep $\Lambda(\alpha)$ real:
\be
e^{-2i\alpha}\in \mathbb{R}
 \Longleftrightarrow 
\mathrm{Re}\,\alpha = \frac{n\pi}{2},\quad n\in\mathbb{Z}.
\ee
If $n$ is odd, there is a sign flip AdS $\leftrightarrow$ dS, and if $b\neq 0$, the magnitude rescales by $e^{2b}$.
Transforming $L\to e^{i\alpha}L$ induces:
\begin{equation}
    \Lambda\to \Lambda(\alpha)=\Lambda_{\text{AdS}}e^{-2i\alpha}.
\end{equation}
Choosing $\alpha=\pi/2$ gives\footnote{We should note that $\alpha$ and $\theta$ are not the same parameter, but they play the same role physically; moving along the analytic AdS–dS family, but they live in two different but equivalent parametrizations. $\theta$ is the phase of $\Lambda$ and $\alpha$ is phase of $L$, and because $\Lambda\sim 1/L^2$, we find $\theta=-2\alpha$}:
\begin{equation}
    \Lambda(\alpha)=\frac{d(d-1)}{2L^2},
\end{equation}
the dS value. This is the same continuation often written as:
\begin{equation}
    L\to \frac{i}{H}, \qquad z\to i\eta,
\end{equation}
the Fefferman-Graham to flat slicing \cite{Skenderis2010, McFaddenSkenderis2010}. 
To make the AdS–dS bridge concrete, we use the ambient hyperboloid embeddings
in:
\begin{equation}
\label{eq:ambient-quadrics}
X^{\!T}\eta_{\mathrm{AdS}}X=-L^2
\quad\longleftrightarrow\quad
X^{\!T}\eta_{\mathrm{dS}}X=+H^{-2},
\qquad (X^4\mapsto iX^4,  L\mapsto i/H),
\end{equation}
and the inner automorphism given by:
\begin{equation}
\label{eq:inner-auto}
\varphi:\ \mathfrak{so}(5,\mathbb{C})\to \mathfrak{so}(5,\mathbb{C}),\quad
\varphi(X)=SXS^{-1},
\end{equation}
where $S=\mathrm{diag}(1,1,1,1,i)$ conjugates
$\mathfrak{so}(3,2)\!\to\!\mathfrak{so}(4,1)$. Equivalently, the curvature
radius is analytically continued as $R_{\mathrm{AdS}}\mapsto i\,R_{\mathrm{dS}}$,
such as $\Lambda\mapsto-\Lambda$, selecting the de Sitter real form without altering the local field content on the $3{+}1$ real slice.
\begin{figure}[H]
  \centering
  \includegraphics[width=1\linewidth]{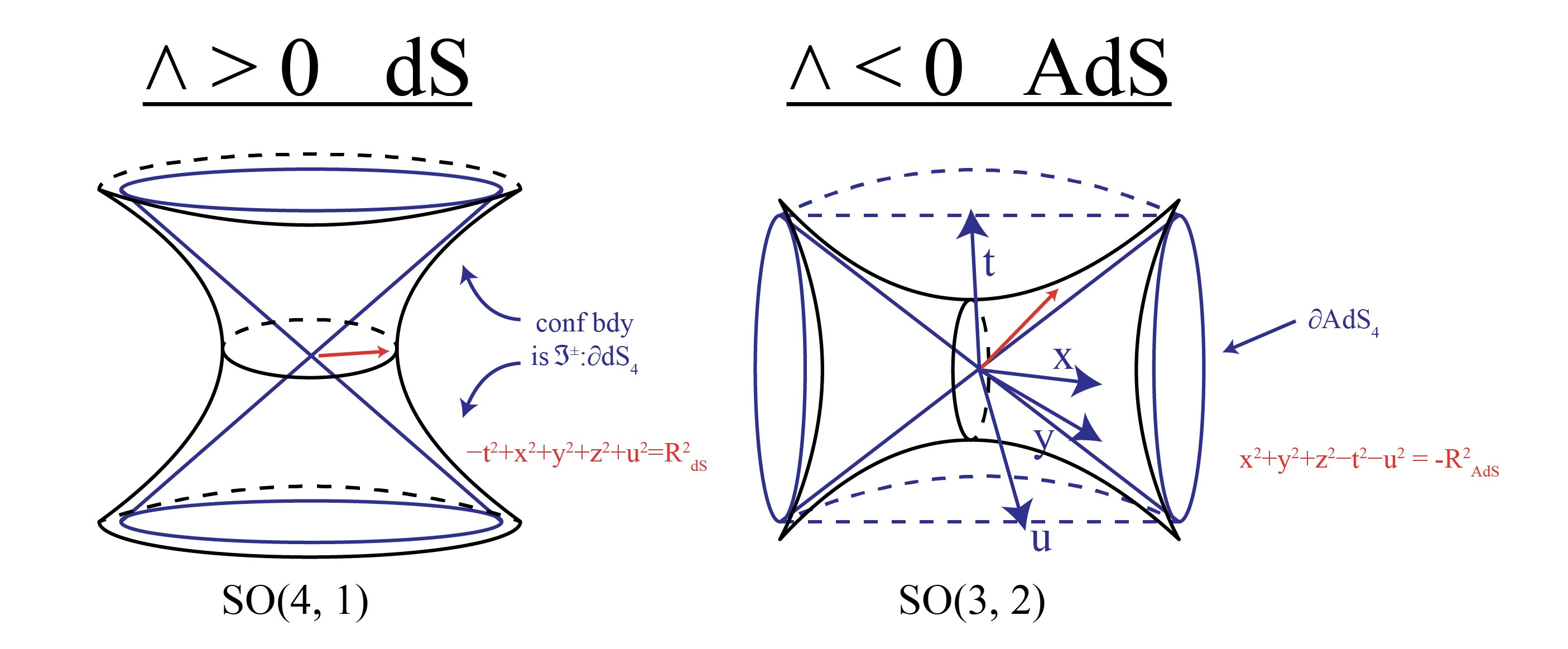}
  \caption{Global embeddings and conformal boundaries of dS\(_4\) and AdS\(_4\).
  The panels show the two real forms SO\((4,1)\) and SO\((3,2)\) of the same
  holomorphic ambient symmetry and the analytic continuation
  \(iR_{\mathrm{dS}}\!\mapsto\! R_{\mathrm{AdS}}\) (\(\Lambda\!\mapsto\!-\Lambda\)) \cite{Strominger2001dSCFT}. The global embeddings and conformal boundaries of dS$_4$ and AdS$_4$.
are $-t^{2}+x^{2}+y^{2}+z^{2}+u^{2}=R_{\mathrm{dS}}^{2}$ in $\mathbb{R}^{1,4}$ (group $\text{SO}(4,1)$) with spacelike boundaries $\Im^\pm \cong S^{3}$ \cite{SpradlinStromingerVolovich2001}.
And $-t^{2}-u^{2}+x^{2}+y^{2}+z^{2}=-R_{\mathrm{AdS}}^{2}$ in $\mathbb{R}^{2,3}$ (group $SO(3,2)$) with timelike boundary $\partial\mathrm{AdS}_4 \cong S^{2}\times\mathbb{R}$.}
  \label{fig:AdS-ds-global}
\end{figure}
Figure~\ref{fig:AdS-ds-global} highlights the causal and topological contrast;
$\partial\mathrm{AdS}_4$ is timelike and on the universal cover diffeomorphic to
$S^2\times\mathbb{R}$ \cite{AvisIshamStorey1978}\footnote{We work with the universal cover of AdS to avoid closed timelike curves \cite{AvisIshamStorey1978, FriedmanSchleichWitt1993, Hawking1992}; on the quotient with time identified the boundary is $S^{2}\!\times\!S^{1}$. Throughout we work with the universal cover $\widetilde{\mathrm{AdS}}_{4}$ and, more generally, restrict the real slice to stably causal hence globally hyperbolic, backgrounds so the \textbf{in–in} formulation is well posed; this avoids the closed timelike curves present on AdS with identified global time. See Friedman–Schleich–Witt on topological censorship and Hawking’s chronology protection; for a recent topology-oriented survey of CTCs see \cite{YoonesyaanRiazi2025}. For dS use ambient the metric, one timelike in the ambient metric (one negative eigenvalue) $\eta=\text{diag}(-,+,+,+,+)$ in $\mathbb{R}^{1,2}$ with coordinates $-X_0^2+X_1^2+X_2^2+X_3^2+X_4^2=R^2_{\text{dS}}$, and for AdS there are two timelike in the ambient metric $\eta=\text{diag}(-,+,+,+,-)$ in $\mathbb{R^{2,3}}$ with coordinates $-X_{-1}^2-X_0^2+X_1^2+X_2^2+X_3^2=R^2_{\text{AdS}}$.}, whereas de Sitter has two spacelike conformal boundaries
$\Im^\pm$ with $\partial\mathrm{dS}_4\simeq S^3\times S^0$ two components. This explains why a unitary boundary Hilbert space exists in AdS but generically fails in dS; our construction therefore computes in AdS, then obtains the late–time dS wavefunctional by analytic continuation, while bulk evolution on the real slice
is handled in the \textbf{in–in} formalism.

Crucially, the continuation changes only the background, the sign of $\Lambda$
and the curvature radius, not the local field content, on the real slice we
retain the same Diff$\times G$ dynamics for gravity and the gauge or matter sector.
All scalar curvature invariants remain polynomial and finite along the continuation,
so there is no geometric singularity obstructing the AdS to dS map.

In HUFT, the bulk action is holomorphic on $M_\mathbb{C}$ and the path integral is taken on a contour $C$ homologous to the real slice. The $\Lambda$ term contributes to the gravitational action in $D=d+1$ spacetime dimensions:
\begin{equation}
    S_{\Lambda}=\frac{1}{\kappa}\int_C d^{d+1}z\sqrt{-\text{det}g_{(\mu\nu)}}\Lambda(\theta),
\end{equation}
where $\kappa$ is the gravitational coupling. If $\Lambda(\theta)$ has a phase, then $e^{iS}$ acquires a damping or phase from $\Im\Lambda$ that changes which Lefschetz thimbles are picked out. As we vary $\theta$, you cross Stokes lines where the dominant saddle can give AdS thimble to dS thimble \cite{Witten2011CS,TanizakiKoike2014}.
The Gibbons-Hawking-York boundary term on the real slice:
\begin{equation}
S_{\text{GHY}}=\frac{\varepsilon}{\kappa}\int_{\partial M} \sqrt{|\gamma|}K, \qquad \varepsilon n_\mu n^\mu =\begin{cases}
    +1, & \text{timelike  boundary (AdS)},\\[4pt]
-1, & \text{spacelike boundary (dS)},
    \end{cases}
\end{equation}
flips sign because the causal character of the cutoff surface flips, while Brown–York variational formulas remain unchanged \cite{York1972,BrownYork1993,BalasubramanianKraus1999}. Where $M$ is the real slice bulk spacetime, $\partial M$ is its boundary such as a timelike conformal boundary of AdS, or spacelike past or future boundaries of dS, $n^\mu$ is the unit outward-pointing normal to $\partial M$; if $\partial M$ is timelike $n^\mu$ is spacelike, so $n_\mu n^\mu = +1$; if $\partial M$ is spacelike, the unit outward normal $n^\mu$ is timelike, so $n_\mu n^\mu = -1$; $\varepsilon$ is the sign parameter that records whether the boundary is timelike or spacelike;
$\varepsilon\in\{+1,-1\}$ is chosen so that $\varepsilon\, n_\mu n^\mu=\pm 1$ as above, this unifies conventions and yields the correct overall sign of the boundary term in Lorentzian signature. $\gamma_{\mu\nu}$ is the induced metric on $\partial M$:
  \begin{equation}
    \gamma_{\mu\nu} = g_{\mu\nu} - \varepsilon\, n_\mu n_\nu,
  \end{equation}
with determinant $\gamma$; the measure uses $\sqrt{|\gamma|}$ to handle both time and spacelike boundaries, $K_{\mu\nu}$ is the extrinsic curvature of $\partial M$:
  \begin{equation}
    K_{\mu\nu}
    = \gamma_{\mu}^{\ \alpha}\,\gamma_{\nu}^{\ \beta}\,\nabla_{\alpha} n_{\beta},
    \qquad
    K = \gamma^{\mu\nu} K_{\mu\nu}\, ,
  \end{equation}
such as $K$ is the trace of $K_{\mu\nu}$. The sign of $K$ depends on the normal orientation; the choice of $\varepsilon$ above ensures a well-posed Dirichlet variational problem for fixed $\gamma_{\mu\nu}$.
On piecewise-smooth boundaries with corners or joints, include Hayward joint terms so for null boundaries, use the null-boundary analogue instead of the GHY form above.

Under the $\theta$-rotation that transforms AdS to dS, the causal character of the cutoff surface flips, so $\varepsilon$ flips sign, and the holographic counterterms built from $\gamma_{ij}$ pick up the corresponding sign or phase. The variational formulas, such as the Brown-York stress tensor or current, are otherwise identical.

\section{Group Level Mapping of Spacetime}

HUFT packages spacetime as either, AdS or dS and the internal gauge sector into a single holomorphic symmetry and connection, then shows how ordinary Einstein–$\Lambda$ gravity and Yang–Mills re-emerge on the real slice and why this setup never violates the Coleman–Mandula theorem in the flat S-matrix regime. A naïve bosonic grand group that entangles spacetime and internal generators would clash with the Coleman–Mandula theorem in flat space but HUFT never violates this as will be shown. We want AdS$_4$ and dS$_4$ to be two real forms of one ambient structure, selected by the $\Lambda$-phase or contour, not two unrelated theories. We as well show explicitly that a single (A)dS connection $(\omega,e)$ reproduces Einstein–$\Lambda$ dynamics through MacDowell–Mansouri formalism. HUFT is formulated on a complexified ambient manifold $M_\mathbb{C}$ with a holomorphic symmetry\footnote{HUFT’s unified group is a statement about the off-shell or ambient symmetry and the geometry, and about AdS/dS where no S-matrix exists. When we ask for on-shell scattering in flat space, the symmetry reduces to the direct product or its supersymmetric graded extension, in line with the Coleman–Mandula theorem.}:
\begin{equation}
    \mathcal{H}_\mathbb{C}= \text{SO}(5, \mathbb{C})\times \mathcal{G}_\mathbb{C},
\end{equation}
where $\mathcal{G}_\mathbb{C}$ is the complexified internal gauge group.  In this symmetry group, it contains SO($3,2$) the AdS$_4$ isometries and SO($4,1$) the dS$_4$ isometries, and $\mathcal{G}_\mathbb{C}$ is the complexified internal gauge group such as $SU(3)\times SU(2)\times U(1)$ or a GUT group \cite{Schrodinger1950,Rindler1960,ParikhSavonijeVerlinde2003}. A real slice is picked by the integration contour or the phase of $\Lambda$, this selects a real form:
\begin{equation}
    \theta=\pi\implies \text{SO}(3,2)\times \mathcal{G}, \qquad \theta=0\implies \text{SO}(4,1) \times \mathcal{G}.
\end{equation}
We treat gravity itself as a gauge theory of the AdS or dS group, and introduce a single complex principal connection:
\begin{equation}
    \mathbb{A}=\frac{1}{2}\mathcal{A}^{AB}J_{AB} \space\oplus\space A^aT_a \in \mathfrak{so}(5, \mathbb{C})\space\oplus\space \mathfrak{g}_\mathbb{c},
\end{equation}
where $\mathbb{A}$ is the unified connection 1-form, $\mathcal{A}^{AB}$ are components of the SO$(5,\mathbb{C})$-valued part of the connection, $A^a$ is the internal gauge connection 1-form valued, and $T_a$ are generators of the internal Lie algebra. With curvature:
\begin{equation}
 \mathbb{F}=d\mathbb{A}+d\mathbb{A}\wedge\mathbb{A}=\frac{1}{2}\mathcal{F}^{AB}J_{AB}\space\oplus\space F^aT_a,
\end{equation}
where $\mathbb{F}$ is the unified curvature 2-form, the field strength of the holomorphic connection, $d$ is the exterior derivative on spacetime, $\mathcal{F}^{AB}$ are curvature components in the $\mathfrak{so}(5,\mathbb{C})$ sector, and $F^a$ is curvature in the internal gauge sector.
We split $A,B=(a,4)$ with $a=1...3$ and define on the real slice:
\begin{equation}
    \mathcal{A}^{ab}=\omega^{ab}, \qquad \mathcal{A}^{a4}=\frac{1}{L}e^a,
\end{equation}
with $\omega^{ab}$ the spin connection, $e^a$ the vierbein, and $L$ the curvature radius; so that:
\begin{equation}
    \mathcal{F}^{ab}=R^{ab}(\omega) \mp \frac{1}{L^2}e^a \wedge e^b, \qquad \mathcal{F}^{a4}=\frac{1}{L}T^a=\frac{(De)^a}{L},
\end{equation}
where $\mathcal{F}^{ab}$ is the curvature 2-form in the (A)dS/Lorentz block, $R^{ab}(\omega)$ is the Lorentz curvature 2-form, $\mathcal{F}^{a4}$ the mixed components encode torsion, and $(De)^a$ the Lorentz-covariant exterior derivative of the coframe, with the upper and lower signs for AdS dS after $L\to i/H$. This repackages Einstein-$\Lambda$ gravity into a single (A)dS connection. Gravity and gauge live in one ambient Holomorphic group $\mathcal{H}_\mathbb{C}$. While different real slices pick SO($3,2$) or SO($4,1$), while the internal $\mathcal{G}$ is common, such that the same unified connection $\mathbb{A}$ governs both.

Real, physical backgrounds are chosen by an anti-linear involution selected by the $\Lambda$-phase whose fixed subalgebra gives the real form:
\begin{equation}
    \theta=\pi: \mathfrak{h}_{\text{AdS}}= \mathfrak{so}(3,2)\oplus \mathfrak{g}, \qquad \theta=0: \mathfrak{h}_{\text{dS}}= \mathfrak{so}(4,1)\oplus \mathfrak{g},
\end{equation}
Therefore, AdS$_4$ and dS$_4$ appear as two real forms of the same holomorphic structure, $\mathfrak{g}$ is the real internal Lie algebra.

We let $J_{AB}=-J_{BA}$ where $A,B=0,1,2,3,4$ generate $\mathfrak{so}(p,q)$ and $T_a$ generate $\mathfrak{g}$. The direct-product structure means:
\begin{equation}
    [J_{AB},J_{CD}]=i(\eta_{BC}J_{AD}-\eta_{AC}J_{BD}+\eta_{AD}J_{BC}-\eta_{BD}J_{AC}), \qquad [T_a,T_a]=if_{ab}^cT_c, \qquad [J_{AB}, T_{a}]=0,
\end{equation}
where $\eta_{Ab}$ is the invariant bilinear form used to raise or lower indices and appearing in the structure constants of the algebra, $f_{ab}^c$ the structure constants of $\mathfrak{g}$ defined by the internal commutator. The vanishing cross-commutators are the algebraic signature that unification through groups in HUFT is not a mixing of spacetime and internal generators at the Lie-algebra level.

We can describe the gravitational sector as a gauge theory of (A)dS through MacDowell-Mansouri \cite{MacDowellMansouri1977}. We split $A,B=(a,4)$, with $a=0,1,2,3$ and introduce an (A)dS valued connection:
\begin{equation}
    \mathcal{A}=\frac{1}{2}\omega^{ab}J_{ab}+\frac{1}{L}e^aJ_{a4},
\end{equation}
where $J_{ab}$ are the Lorentz generators, and $J_{a4}$ the (A)dS transvection generators, with vierbein $e^a$ and spin connection $\omega^{ab}=-\omega^{ba}$. Its curvature is:
\begin{equation}
    \mathcal{F}=d\mathcal{A}+\mathcal{A}\wedge\mathcal{A}=\frac{1}{2}\mathcal{F}^{ab}J_{ab}+\mathcal{F}^{a4}J_{a4},
\end{equation}
where:
\begin{equation}
    \mathcal{F}^{ab}=R^{ab}(\omega)\mp\frac{1}{L^2}e^a\wedge e^b, \qquad \mathcal{F}^{a4}=\frac{1}{L}T^a:=\frac{1}{L}(De)^a,
    \label{curvatureF}
\end{equation}
upper signs for AdS and lower for dS, after $L\to i/H$. The MacDowell-Mansouri action gives Einstein-$\Lambda$ plus a topological term, we consider:
\begin{equation}
    S_{\text{MM}}=\frac{1}{2\kappa}\int\epsilon_{abcd} \mathcal{F}^{ab}\wedge\mathcal{F}^{cd},
    \label{MacDowell}
\end{equation}
where $\epsilon_{abcd}$ is the totally antisymmetric Levi-Civita symbol in the internal Lorentz frame, tangent-space indices; it is the invariant tensor that projects the SO(3,2) or SO(4,1) curvature onto Lorentz pieces and yields the top-form needed for the integral. Expanding using \ref{curvatureF}:
\begin{equation}
\epsilon_{abcd}\mathcal{F}^{ab}\wedge\mathcal{F}^{cd}=\epsilon_{abcd}(R^{ab}\mp\frac{1}{L^2}e^a\wedge e^b)\wedge(R^{cd}\mp\frac{1}{L^2}e^c\wedge e^d),
\end{equation}
and hence:
\begin{equation}
    \epsilon_{abcd}\mathcal{F}^{ab}\wedge\mathcal{F}^{cd}= \underbrace{\epsilon_{abcd}R^{ab}\wedge R^{cd}}_{\text{Gauss-Bonnet (topological in 4D)}} \mp \frac{2}{L^2}\epsilon_{abcd}R^{ab}\wedge e^c\wedge e^d+\frac{1}{L^4}\epsilon_{abcd}e^a\wedge e^b \wedge e^c \wedge e^d,
\end{equation}
thus:
\begin{equation}
    S_{\text{MM}}=\frac{1}{2\kappa}\int\epsilon_{abcd}(R^{ab}\wedge e^c\wedge e^d \mp \frac{1}{L^2}e^a \wedge e^b \wedge e^c \wedge e^d)+S_{\text{GB}},
\end{equation}
where $S_{\text{GB}}$ is topological in 4D. Identifying $\Lambda=\mp3/L^2$ and $\epsilon_{abcd}R^{ab}\wedge e^c\equiv2 \text{vol}\times R(h)$ gives the Einstein-Hilbert term plus a cosmological constant. Variation with respect to $e^a$ and $\omega^{ab}$, setting torsion $T^a=0$ gives the Einstein field equations.

We can add the internal gauge and matter sectors. We let $A=A^aT_a$ be the internal connection with curvature $F=dA+A\wedge A$, the standard Yang-Mills pieces:
\begin{equation}
    S_{\text{YM}}=-\frac{1}{4g_{\text{YM}}}\int\tr(F\wedge*F),
\end{equation}
where $g_{\text{YM}}$ is the Yang–Mills gauge coupling, $F$ is the internal field-strength 2-form of the gauge connection, and $*F$ the spacetime Hodge dual of $F$ with respect to the metric used in the matter or gauge sector.
Varying $S_{\text{YM}}+S_{\text{matter}}$ adds:
\begin{equation}
    T_{\mu\nu}^{\text{YM}}=F_{\mu\alpha}F_\nu^\alpha-\frac{1}{4}h_{\mu\nu}F_{\alpha\beta}F^{\alpha\beta}, \qquad D_\mu F^{\mu\nu}=J^\nu,
\end{equation}
to the gravitational equation. The total action is:
\begin{equation}
    S=S_{\text{MM}}[e,\omega]+S_{\text{YM}}[A;h]+S_{\text{matter}}[h,A,\Psi],
\end{equation}
one geometric package for gravity with $\Lambda$, and the internal gauge sector $A$, all couched in a single ambient symmetry but with commuting subalgebras.

The Coleman-Mandula theorem requires Poincar\'e invariance of the S-matrix \cite{ColemanMandula1967,HaagLopuszanskiSohnius1975}:
\begin{equation}
    [P_\mu, S]=[M_{\mu\nu},S]=0,
\end{equation}
where $P_\mu$ and $M_{\mu\nu}$ are generators of the Poincar\'e group, translations and Lorentz transformations, and $S$ is the scattering operator. Nontrivial scattering, some $2\to2$ amplitudes are not identically zero and not concentrated only at forward/backward angles. Analyticity and polynomial boundedness of amplitudes in the usual Mandelstam variables $(s,t,\ldots)$, microcausality plus Lehmann–Symanzik–Zimmermann (LSZ) reduction. Cluster decomposition: distant experiments decouple. Finite particle spectrum growth, for any mass $m$, only finitely many species with mass $\leq m$ and finite spin. Global, not gauge continuous symmetries acting on one-particle states by finite-dimensional rep’s. These are standard S-matrix conditions in $D\geq3$. It rules out a grand bosonic group that intertwines spacetime and internal symmetries, like a naive SO$(4,2+n)$ acting linearly on a single multiplet of particles in flat space. It explains why in the real world we see Poincar\'e × (internal), with internal possibly spontaneously broken, rather than a bigger simple bosonic group.

We can In\"on\"u-Wigner contract to flat spacetime to check that the Coleman-Mandula theorem is not violated \cite{InonuWigner1953}. To discuss the Coleman-Mandula theorem we must move to a spacetime where it applies, a regime with a Poincar\'e invariant S-matrix \cite{ColemanMandula1967,HaagLopuszanskiSohnius1975}. This is obtained by contracting $\mathfrak{so}(3,2)$ or $\mathfrak{so}(4,1)$ to the Poincar\'e algebra $\mathfrak{p}$ by sending the curvature scale to infinity for AdS $L\to\infty$ or to zero for dS $H\to0$. We define for AdS:
\begin{equation}
    M_{\mu\nu}:=J_{\mu\nu}, \qquad P_{\mu}:=\frac{1}{L}J_{\mu4}, \qquad \mu,\nu=0,1,2,3.
\end{equation}
Using the $\mathfrak{so}(3,2)$ commutators we find:
\begin{equation}
    [M_{\mu\nu}, M_{\rho\sigma}]=i(\eta_{\nu\rho}M_{\mu\sigma}-\eta_{\mu\rho}M_{\nu\sigma}+\eta_{\mu\sigma}M_{\nu\rho}-\eta_{\nu\sigma}M_{\mu\rho}),
\end{equation}
\begin{equation}
    [M_{\mu\nu},P_\rho]=i(\eta_{\nu\rho}P_\mu-\eta_{\mu\rho}P_\nu),
\end{equation}
\begin{equation}
    [P_\mu,P_\rho]=\pm\frac{i}{L^2}M_{\mu\rho}\underset{L\to\infty}{\longrightarrow} 0,
\end{equation}
where the sign corresponds to AdS or dS. in the flat limit $L\to\infty$ or $H\to0$, these become the Poincar\'e brackets:
\begin{equation}
    [M,M]\sim\mathfrak{so}(3,1), \qquad [M,P]\sim P, \qquad [P,P]=0.
\end{equation}
The total algebra before contraction is a direct sum:
\begin{equation}
    \mathfrak{h=(A)ds\oplus g}, \qquad [J_{AB}, T_a]=0,
\end{equation}
\begin{equation}
    \mathfrak{h} = \mathfrak{so}(3,2) \oplus \mathfrak{g}
\quad\text{or}\quad
\mathfrak{h} = \mathfrak{so}(4,1) \oplus \mathfrak{g}.
\end{equation}
Therefore, after contraction:
\begin{equation}
    [P_\mu,\mathcal{Q}^a]=0, \qquad [M_{\mu\nu}, \mathcal{Q}^a]=0,
\end{equation}
for the conserved charges $\mathcal{Q}^a$. The flat-space symmetry is then:
\begin{equation}
    \mathfrak{h}_{\text{flat}}=\mathfrak{p\oplus g},
\end{equation}
with no non-trivial bosonic mixing. So, in the S-matrix regime, HUFT's symmetry is exactly the Coleman-Mandula allowed direct product. 
In this holomorphic setting, gravity and gauge fields are encoded by a single ambient symmetry and connection; AdS and dS arise as distinct real forms selected by the $\Lambda$-phase, while the internal $G$ sector is common to both. The commuting subalgebras guarantee that no bosonic spacetime–internal mixing is introduced, and in the flat limit the symmetry contracts to $\mathfrak{p}\oplus \mathfrak{g}$ as required by Coleman–Mandula. This group-level packaging therefore provides a mathematically clean bridge between AdS and dS that reproduces Einstein–$\Lambda$ dynamics and preserves S-matrix consistency where applicable.

\section{AdS to dS Singularity Evaluation}
While one may think there is a global singularity present when $\Lambda=0$ there is in fact not as the phase space formulation forbids it as any time when $\Lambda=0$ is purely complex, so there is no singularity forced when moving between AdS and dS through HUFT's analytic continuation when it is done through the ambient complex manifold $M_\mathbb{C}$. The continuation is non-singular as the constant curvature vacua are analytic in $\Lambda$. In $D=d+1$ dimensions, a maximally symmetric metric has:
\begin{equation}
    R_{\mu\nu\rho\sigma}=K(g_{\mu\rho}g_{\nu\sigma}-g_{\mu\sigma}g_{\nu\rho}), \qquad R_{\mu\nu}=K(D-1)g_{\mu\nu}, \qquad R=KD(D-1),
\end{equation}
where $R_{\mu\nu\rho\sigma}$ is the Riemann curvature tensor,  $R_{\mu\nu}$ is the Ricci tensor, $R$ is the Ricci scalar, and $K$ is the constant sectional curvature. We insert $R_{\mu\nu}=Kg_{\mu\nu}$ and $R=D(D-1)K$ into the Einstein-$\Lambda$ equations:
\begin{equation}
    R_{\mu\nu}-\frac{1}{2}Rg_{\mu\nu}+\Lambda g_{\mu\nu}=0,
\end{equation}
resulting in:
\begin{equation}
    (D-1)K\frac{1}{2}D(D-1)K+\Lambda=0 \implies K=\frac{2\Lambda}{(D-1)(D-2)}.
\end{equation}
This shows that all curvature invariants are polynomials in $\Lambda$. Some useful definitions are:
\begin{equation}
    R=\frac{2D}{D-2}\Lambda, \qquad R_{\mu\nu}R^{\mu\nu}=\frac{4D}{(D-2)^2}, \qquad R_{\mu\nu\rho\sigma}R^{\mu\nu\rho\sigma}=\frac{8D(D-1)}{(D-2)^2(D-1)^2}\Lambda^2 \propto \Lambda^2.
\end{equation}
We can use this to conclude that flipping between $\Lambda\to-\Lambda$ leaves all invariants finite and unchanged when they are even in $\Lambda$.

The explicit AdS to dS map has no singular Jacobians, we start with $\text{AdS}_D$ in Fefferman-Graham or Poincar\'e form \cite{Skenderis2010, McFaddenSkenderis2010}:
\begin{equation}
    ds^2_{\text{AdS}}=\frac{L^2}{z^2}(dz^2+\eta_{ij}dx^idx^j), \qquad z>0, \qquad \Lambda_{\text{AdS}}=-\frac{(D-1)(D-2)}{2L^2},
\end{equation}
where $\eta_{ij}$ is the flat (D–1)-dimensional Minkowski metric on the boundary coordinates $x^i$ of the AdS Poincar\'e patch, and $z$ the AdS radial coordinate. When we make the analytic continuation:
\begin{equation}
    z\to i\eta, \qquad L\to \frac{i}{H},
\end{equation}
and writing $\eta<0$. Then we find:
\begin{equation}
    \frac{L^2}{z^2}(dz^2+\eta_{ij}dx^idx^j)\to \frac{1}{(H\eta)^2}(-d\eta^2+\eta_{ij}dx^idx^j)=ds^2_{\text{dS}},
\end{equation}
which is the flat slicing of $dS_D$ with:
\begin{equation}
    \Lambda_{\text{dS}}=+\frac{(D-1)(D-2)}{2}H^2.
\end{equation}
We see the Jacobian of the map is finite and non-zero, it is just phases of $i$, and no component of the metric blows-up away from the usual conformal boundary blows up $z\to 0 \quad \text{or} \quad \eta\to0^-$, which is not a curvature singularity it is a conformal singularity, as all invariants remain constant.

The extrinsic geometry shows this as the Gibbons-Hawking-York data stays finite. The induced metric at the cutoff:
\begin{equation}
    \sqrt{|\gamma|}_{\text{AdS}}=\left(\frac{L}{z}\right)^d, \qquad \sqrt{|\gamma|}_{\text{dS}}=\left(
    \frac{1}{H|\eta|}\right)^d.
\end{equation}
The unit normals outward:
\begin{equation}
    n^\mu_{\text{AdS}}=\frac{z}{L}\delta^\mu_z \qquad \text{(timelike boundary,}\quad n^2=+1), \qquad n_{\text{dS}}^\mu=-H\eta\delta_\eta^\mu, \quad (\text{spacelike screen}, \quad n^2=-1).
\end{equation}
The extrinsic curvatures:
\begin{equation}
    K_{\text{AdS}}=\nabla_\mu n^\mu=-\frac{d}{L}, \qquad K_{\text{dS}}=\nabla_\mu n^\mu=+dH.
\end{equation}
Both are finite and map to each other under $L\to1/H$ with a sign from the causal type of the boundary. The Gibbons-Hawking-York term is finite on both sides; no boundary pathology arises from the continuation.

In D=4, the static and sphericity symmetric vacuum with $\Lambda$:
\begin{equation}
R_{\mu\nu\rho\sigma}R^{\mu\nu\rho\sigma}=C_{{\mu\nu\rho\sigma}}C^{\mu\nu\rho\sigma}+\frac{1}{6}R^2=\underbrace{\frac{48G^2M^2}{r^6}}_{\text{Weyl}^2}+\underbrace{\frac{8}{3}\Lambda^2}_{\text{Constant curvature}}, 
\end{equation}
where $R_{\mu\nu\rho\sigma}R^{\mu\nu\rho\sigma}$ is the Kretschmann scalar, $C_{\mu\nu\rho\sigma}$ the Weyl tensor.
The only true curvature singularity is at $r=0$. The horizons are the real positive roots of $f_{\Lambda}(r)=0$. For dS:
\begin{equation}
    H^2\equiv\frac{\Lambda}{3} \implies H^2r^3-r+2GM=0.
\end{equation}
The discriminant:
\begin{equation}
    \Delta=\frac{1}{H^4}\left(\frac{1}{27}-G^2M^2\Lambda\right),
\end{equation}
where $\Delta$ is the scaled discriminant of the cubic horizon equation for Schwarzschild–de Sitter, which determines the horizon structure.
For $0<9G^2M^2\Lambda<1$, there are two positive roots $r_b<r_c$ for black holes and cosmological horizons. At the Nariai limit $9G^2M^2\Lambda\to1$, they merge \cite{Nariai1951}:
\begin{equation}
    r_b=r_c=\frac{1}{\sqrt{\Lambda}}.
\end{equation}
Curvature invariants at the degenerate horizon are finite:
\begin{equation}
    R=4\Lambda, \qquad R_{\mu\nu}R^{\mu\nu}=4\Lambda^2, \qquad K=\frac{8}{3}\Lambda^2+\frac{48G^2M^2}{r^6}\Bigg|_r=1/\sqrt{\Lambda}<\infty.
\end{equation}
For AdS:
\begin{equation}
    f_\Lambda(r)=1-\frac{2GM}{r}+\frac{r^2}{L^2}, \qquad (\text{with}) \quad L^2=-\frac{3}{\Lambda},
\end{equation}
there is no single positive horizon for $M>0$, as $\Lambda$ crosses zero, the real roots vary continuously; no divergences occur in the metric coefficients or invariant scalars.

The boundary generating function $W$ is nonsingular under the transition between AdS to dS. We let $W_{\text{AdS}}[\gamma, A,...]$ be the renormalized on shell action, Euclidean AdS with $W_{\text{dS}}=\text{In}\Psi_{\text{dS}}$ as the late time wavefunctional, a Lorentzian in dS. We evaluate the bulk action and the Gibbons-Hawking-York term on a cutoff surface and remove the cutoff with local counterterms. Under the analytic continuation, we find at quadratic order in sources and similarly to all orders diagrammatically. We start from the AdS expression written in terms of the AdS radius \(L\):
\begin{equation}
W_{\text{AdS}} = W_{\text{AdS}}(L).
\end{equation}
To go to de Sitter we perform the holomorphic continuation $L  \longrightarrow  \frac{i}{H}$. So naively we would write:
\begin{equation}
W_{\text{dS}} \stackrel{?}{=} W_{\text{AdS}}\bigl(L = i/H\bigr).
\end{equation}
Now suppose that \(W_{\text{AdS}}\) contains an overall scaling with the curvature radius of the form:
\begin{equation}
W_{\text{AdS}}(L)  \propto  L^{-(d-1)} \, \mathcal{F},
\end{equation}
where \(\mathcal{F}\) is real after counterterms and \(d\) is the boundary spacetime dimension. Then substituting \(L = i/H\) gives:
\begin{equation}
W_{\text{AdS}}\bigl(L = i/H\bigr)
 = 
\bigl(i/H\bigr)^{-(d-1)} \, \mathcal{F}
 = 
i^{-(d-1)} H^{\,d-1} \, \mathcal{F}.
\end{equation}
Equivalently, we can rewrite this as:
\begin{equation}
W_{\text{AdS}}\bigl(L = i/H\bigr)
 = 
i^{\,d-1} \, W_{\text{AdS}}\bigl(L = 1/H\bigr),
\end{equation}
because:
\begin{equation}
W_{\text{AdS}}\bigl(L = 1/H\bigr)
 = 
(1/H)^{-(d-1)} \mathcal{F}
 = 
H^{\,d-1} \mathcal{F}.
\end{equation}
Putting it together, the de Sitter functional can therefore be written as:
\[
W_{\text{dS}}
 = 
i^{\,d-1} \, W_{\text{AdS}}\bigl(L = 1/H\bigr)
 +  (\text{local counterterms}),
\]
which is the form used in the paper: the phase \(i^{\,d-1}\) carries the effect of the complex rotation \(L \to i/H\), and the remaining substitution is made with the real radius \(L = 1/H\) giving\footnote{Throughout the analytic continuation from AdS to dS we formally take the AdS radius \(L \to i/H\), which flips the sign of the cosmological constant. At the level of the on–shell AdS functional, however, it is convenient to factor out the pure phase generated by this rotation, yielding the relation \(W_{\text{dS}} = i^{\,d-1} W_{\text{AdS}}\big|_{L \to 1/H} + (\text{local})\). In other words, the imaginary unit from \(L \to i/H\) is accounted for explicitly in the overall prefactor \(i^{\,d-1}\), so the remaining substitution in the functional is made with a real radius \(L = 1/H\).}:
\begin{equation}
    W_{\text{dS}}[\text{sources}]=i^{d-1}W_{\text{AdS}}[\text{sources}]\bigg|_{L\to 1/H}+\text{local counterterms},
\end{equation}
consistent with explicit late-time wavefunctional calculations \cite{Skenderis2002,MaldacenaPimentel2011}.
The factor of $i^{d-1}$ is the product of, one $i$ from $dz\to id\eta$, $d$ powers of $i$ from $\sqrt{|\gamma|}\sim z^{-d}\to (i\eta)^{-d}$, and a sign from the change of the unit normal's causal type in the Gibbons-Hawking-York term. This mapping introduces no new poles and leaves all local counterterms analytic in the curvature scale \cite{Moffat:FiniteNonlocal1990,Tomboulis1997,Biswas2012,Talaganis2015,Buoninfante2018,MT:FiniteHolomorphicQFT}. This means that the holographic 1-pt functions:
\begin{equation}
    \langle T_{ij}\rangle=\frac{2}{\sqrt{\gamma}}\frac{\delta W}{\delta\gamma^{ij}}, \qquad \langle J^i\rangle=\frac{1}{\sqrt{|\gamma|}}\frac{\delta W}{\delta A_i},
\end{equation}
exist and remain finite on both sides, the only change is the overall phase in even or odd $d$ that is responsible for the lack of reflection positivity on the dS screen.

In HUFT, edges can appear, but they are not singularities. When $\Lambda=0$, this is a flat limit, not a blow-up; some AdS/dS normalizations scale like powers of $L$ or $H^{-1}$, so one must take the flat-space limit with the standard rescaling, such as the Penedones map for S-matrix extraction. The Stokes phenomena in the complex plane, along the $\Lambda$-rotation, which saddle dominates, can change the Stokes lines \cite{Witten2011CS,TanizakiKoike2014}, but that is a change of dominant thimble, not a geometric singularity. The endpoints at $\theta=\pi$ and $\theta=0$ for AdS and dS are regular. The transition between AdS and dS is a holomorphic rotation of the curvature scale inside $M_\mathbb{C}$. On the real slice, it lands on two regular constant-curvature spacetimes. No new singularity is created by analytic continuation; any singularities that show up are the usual ones, such as $r=0$ for black holes, but these exist regardless of the AdS to dS flip.

\section{Unitarity of the Spacetimes}
In HUFT, we find that bulk unitarity holds for both AdS and dS, but boundary unitarity holds in AdS but not in dS. As we will show, this is because for bulk unitarity, we have quantization by a holomorphic path integral on the complexified field space. Since the integration cycle $C$ is chosen by Picard-Lefschetz, the dominant saddle is the real slice $y^\mu=0$. Expanding around the saddle that gives ordinary Lorentzian QFT with canonical signs, ghosts ensure gauge and diffeomorphism invariance, and unitarity reduces to the choice of steepest descent contour that keeps the real slice dominant. We work on a ($d+1$)-dimensional Lorentzian manifold $M$ with boundary $\partial M$. The real slice action is:
\begin{equation}
    S[h,A,\Phi]=\frac{1}{2\kappa}\int_M\sqrt{|h|}(R-2\Lambda)-\frac{1}{4g_{\text{YM}}^2}\int_MF_{\mu\nu}F^{\mu\nu}+S_{\text{matter}}[\Phi;h,A]+\frac{\varepsilon}{\kappa}\int_{\partial M}\sqrt{|\gamma|}K,
\end{equation}
where $\varepsilon$ is the sign accounting for the boundary’s causal type $\varepsilon=n_\mu n^\mu=+1$ for a timeline boundary and $-1$ for a spacelike screen, $\kappa$ is the gravitational coupling, $\Phi$ is the matter fields, $\gamma_{ij}$ is the metric induced by $h$ on $\partial M$, $K=\gamma^{ij}K_{ij}$ where $K_{ij}$ is the extrinsic curvature of $\partial M$, $|h|$ is the absolute value of $\det(h_{\mu\nu})$, and $g^2_{\text{YM}}$ is the Yang–Mills coupling. As shown above, when we vary with respect to the symmetric part of the metric $h$, we get the Einstein field equations, and when we vary with respect to the antisymmetric part, we get the gauge Yang-Mills and Maxwell equations.

If we vary the renormalized boundary function $W$ with respect to sources:
\begin{equation}
    \langle T_{ij}\rangle=\frac{2}{\sqrt{\gamma}}\frac{\delta W}{\delta\gamma^{ij}}, \qquad \langle J^i\rangle=\frac{1}{\sqrt{|\gamma|}}\frac{\delta W}{\delta A_i}, \qquad \langle\mathcal{O_\alpha}\rangle=\frac{1}{\sqrt{|\gamma|}}\frac{\delta W}{\delta \phi_{(0)}^\alpha},
\end{equation}
where $W$ is the boundary generating functional, $\gamma_{ij}$ is the induced metric on the boundary, $A_i$ is the boundary value of the bulk gauge field, $\phi^\alpha_{(0)}$ is the boundary source; the leading or non-normalizable mode for a bulk scalar $\phi^\alpha$. With $W=S_{\text{ren}} (\text{AdS})$ or $W=\text{In}\Psi_{\text{dS}}(\text{dS})$. Diffeomorphism and gauge invariance of $W$ give the Ward identities:
\begin{equation}
    \nabla_i\langle T^{ij}\rangle=F^i_j\langle J^i\rangle+\sum_\alpha\langle\mathcal{O_\alpha}\rangle\nabla^j\phi^\alpha_{(0)}, \qquad D_i\langle J^i\rangle=0.
\end{equation}
The trace identity for the Weyl anomaly in even $d$ can be derived as under an infinitesimal Weyl rescaling with parameter $\sigma(x)$, where $\sigma$ is the local Weyl parameter, a smooth, dimensionless function on the boundary, $\sigma=\sigma(x)$, that generates an infinitesimal conformal rescaling of the boundary fields \cite{HenningsonSkenderis1998, DeHaroSkenderisSolodukhin2001}:
\begin{equation}
    \delta_\sigma\gamma_{ij}=2\sigma\gamma_{ij}, \qquad \delta_\sigma A_i=0 \qquad (\text{for a conserved current of dimension} \space d-1), \quad \delta_\sigma\phi_{(0)}^\alpha=(\Delta_\alpha-d)\sigma\phi_{(0)}^\alpha,
\end{equation}
the variation of $W$ is:
\begin{equation}
    \delta_\sigma W=\int d^4x \sqrt{|\gamma|}\sigma(x)(\langle T^i_i\rangle-\sum_\alpha(d-\Delta_\alpha)\phi_{(0)}^\alpha\langle\mathcal{O_\alpha}\rangle).
\end{equation}
By definition, the Weyl anomaly appears as a local functional $A[\gamma, A,...]$\footnote{This is a local scalar density of Weyl weight $d$ built from the boundary sources. It only exists in even $d$.}, on the right:
\begin{equation}
    \delta_\sigma W=\int d^dx\sqrt{|\gamma|}\sigma(x)A[\gamma,A,...].
\end{equation}
Equating the two expressions gives the trace identity:
\begin{equation}
    \langle T^i_i\rangle=A[\gamma,A,...]+\sum_\alpha(d-\Delta_\alpha)\phi_{(0)}^\alpha\langle\mathcal{O_\alpha}\rangle).
\end{equation}
If one also turns on exactly marginal couplings $g^I(x)$ with beta function $\beta^I$, their Weyl variation adds the standard term $\sum_I\beta^I\langle\mathcal{O_I}\rangle$, to the right-hand side.

At the level of the bulk for AdS and dS, HUFT's functional integral is the standard real-time QFT plus exponentially small thimble corrections. To prove bulk unitarity, we chose the Schwinger-Keldysh (CPT) contour $C$. We quantize the real slice theory with the Schwinger-Keldysh (CPT) contour $C$ and initial state $\rho_0$. For compactness, suppress all indices and species:
\begin{equation}
    Z[J_+,J_-]=\text{Tr}(U_+(J_+)\rho_0U_-^\dagger(J_-))=\int D\Phi_+D\Phi_-\exp\{iS[\Phi_+]-iS[\Phi_-]+i\int_C(J_+\Phi_+-J_-\Phi_-)\},
\end{equation}
where $\rho_0$ is the initial density matrix, the state at the initial time used to define expectation values in the \textbf{in-in} formalism. It guarantees unitarity and causality of real-time correlators, replaces the missing global S-matrix in dS with a well-posed \textbf{in-in} framework, and ensures the positivity and thermality properties you need for physically meaningful observables.
Since the largest time identity implies unitarity, we set equal sources $J_+=J_-=J$. Path ordering on $C$ and the fact that the backward branch is this inverse evolution give:
\begin{equation}
    Z[J,J]=1,
\end{equation}
for any background. We differentiate twice to get the matrix of two-point functions $G^{ab},\space(a,b=\pm)$:
\begin{equation}
    G^{++}+G^{--}-G^{+-}-G^{-+}=0,
\end{equation}
the Schwinger-Keldysh unitarity identity. Higher derivatives yield diagrammatic cutting rules, such as the optical theorem, which states that the imaginary part of any time-ordered correlator equals the sum over cuts.

The AdS boundary is unitary as the reflection positivity implies Osterwalder-Schrader reconstruction \cite{OsterwalderSchrader1973,Skenderis2002}. We work in Euclidean AdS, a hyperbolic space, and use the Fefferman-Graham gauge:
\begin{equation}
    ds^2=\frac{L^2}{z^2}(dz^2+\delta_{ij}dx^idx^j), \qquad z\to0.
\end{equation}
A scalar example to solve the equation of motion and evaluate the on-shell action. The Euclidean bulk action:
\begin{equation}
    S_E[\phi]=\frac{1}{2}\int\frac{L^{d+1}}{z^{d+1}}\left[(\partial_z\phi)^2+(\partial_i\phi)^2+\frac{m^2L^2}{z^2}\phi^2\right]dzd^dx.
\end{equation}
Fourier transform $\phi(z,k)$ and solve:
\begin{equation}
    [z^2\partial^2_z-(d-1)z\partial_z+z^2k^2-m^2L^2]\phi=0.
\end{equation}
We set $\nu=\sqrt{(d/2)^2+m^2L^2}, \space \Delta=d/2+\nu$, where $m$ is the bulk scalar mass for the field $\phi$, $\nu$ is the dimensionless Bessel index that appears when solving the bulk scalar EOM in AdS, $\Delta$ is the scaling dimension of the boundary operator $\mathcal{O}$ dual to $\phi$; in standard quantization. The regular interior-normalizable solution is:
\begin{equation}
\phi(z,k)=\phi_{(0)}(k)z^{d/2}K_\nu(kz),
\end{equation}
this is the standard bulk scalar solution in Euclidean AdS after Fourier transforming along the boundary, where $k$ is the magnitude of the Euclidean boundary momentum from the Fourier transform, $K_\nu$ is the modified Bessel function of the second kind; this choice is the interior regular or normalizable mode. This formula encodes the correct radial falloffs that set the source and vev, and regularity in the interior via the $K-\nu$ branch. Near the boundary:
\begin{equation}
    \phi(z,k)=\phi_{(0)}(k)z^{d-\Delta}[1+...]+\phi_{(1)}(k)z^{\Delta}[1+,,,], \qquad \phi_{(1)}(k)=\frac{\Gamma(1-\nu)}{2^{2\nu-1}\Gamma(\nu)}k^{2\nu}\phi_{(0)}(k),
\end{equation}
where $\Gamma$ is the Gamma function; $\nu$ appears as the Bessel index from solving the radial EOM, $z$ is the AdS radial coordinate, $\phi_{(1)}(k)$ is the coefficient of the normalizable mode, proportional to the VEV $\langle\mathcal{O}\rangle$ in standard quantization; it’s fixed by regularity in the interior and relates to $\phi_{(0)}$ by the Bessel-function ratio shown. The on-shell action is a boundary term. Using the outward unit normal $n^z=L/z$:
\begin{equation}
    S_E^{\text{on-shell}}=\frac{1}{2}\int\phi\sqrt{\gamma} \space n^\mu\partial_\mu\phi d^dx=\frac{1}{2}\int\frac{d^dk}{(2\pi)^d}\phi_{(0)}(-k)K_\Delta^{\text{AdS}}(k)\phi_{(0)}(k),
\end{equation}
with the positive kernel:
\begin{equation}
    K_\Delta^{\text{AdS}}(k)=c_\Delta k^{2\Delta-d}, \qquad c_\Delta=\frac{L^{d-1}}{2}\frac{\Gamma(\Delta)}{\Gamma(\Delta-\frac{d}{2})}>0,
\end{equation}
after adding the standard local counterterms that remove contact divergences. The connected two-point function of the dual operator $\mathcal{O}$:
\begin{equation}
    \langle\mathcal{O}(k)\mathcal{O}(-k)\rangle=K_\Delta^{\text{AdS}}(k)=c_\Delta k^{2\Delta-d}, \quad\text{(Euclidean)}.
\end{equation}

To show reflection positivity and Osterwalder-Schrader reconstruction, we let $G_E(x-y)$ be the Fourier transform of $K_\Delta^{\text{AdS}}(k)$. For any test function $f$ supported at Euclidean times $x^0>0$:
\begin{equation}
    \int d^dxd^dyf^*(\theta x)G_E(x-y)f(y)=\int\frac{d^dk}{(2\pi)^d}|\tilde{f}(k)|^2K_\Delta^{\text{AdS}}(k)\ge0,
\end{equation}
because $K_\Delta^{\text{AdS}}(k)\geq0$, this implies reflection positivity. Together with Euclidean invariance and regularity, Osterwalder-Schrader reconstruction yields a unitary Lorentzian Hilbert space and Hamiltonian for the boundary theory.

The same structure holds for conserved currents and the stress tensor. Solving their linearized bulk equations and reinserting gives the transverse kernels of the form:
\begin{equation}
    \langle J_iJ_j\rangle=C_Jk^{d-2}\Pi_{ij}, \qquad \langle T_{ij}T_{mm}\rangle=C_Tk^dP_{ij,lm},
\end{equation}
with positive $C_{J,T}\propto L^{d-3}/{g^2}, \quad L^{d-1}/\kappa$, where $C_{J,T}$ are positive theory-dependent normalizations, central charges for the 2-pt sector, $\Pi_{ij}$ the transverse projector, and $k$ the magnitude of the Euclidean boundary momentum, $k\equiv|\textbf{k}|$. Meaning the AdS theory is unitary.

The dS boundary or screen is not unitary. Using flat slicing:
\begin{equation}
    ds^2=a^2(\eta)(-d\eta^2+d\textbf{x}^2), \qquad a(\eta)=-\frac{1}{(H\eta)}, \qquad \eta\in(-\infty,0).
\end{equation}
The action:
\begin{equation}
    S=\frac{1}{2}\int d\eta\space d^dx\space a^{d-1}[\phi^2-(\nabla\phi)^2-a^2m^2\phi^2],
    \label{dS_late_time_action}
\end{equation}
we solve the mode equation with the Bunch-Davies initial condition:
\begin{equation}
    u^{''}_k+\left(k^2-\frac{\nu^2-\frac{1}{4}}{\eta^2}\right)u_k=0, \qquad u_k(\eta)=\frac{\sqrt{\pi}}{2}e^{i(\nu+\frac{1}{2})\frac{\pi}{2}}(-\eta)^\frac{1}{2}H_\nu^{(1)}(-k\nu), \qquad \nu=\sqrt{\frac{d^2}{4}-\frac{m^2}{H^2}},
\end{equation}
where $u_k(\eta)$ is the canonically normalized mode function, $\eta$ is conformal time, $k$ is the comoving wavenumber, $d$ is the number of spatial dimensions, $H_\nu^{(1)}$ is the Hankel function of the first kind; choosing it with the given phase picks the Bunch–Davies vacuum, and $u_k^n$ where $n=','',...$ is the derivative of the mode function with respect to conformal time. We write the field in Fourier space $\phi(\eta,\textbf{k})=\phi_\textbf{k}u_k(\eta)+\text{h.c.}$ and defined the canonical momentum $\pi=a^d-1\phi'$. For a Gaussian late-time wavefunctional:
\begin{equation}
    \Psi_{\text{dS}}[\varphi]\propto\exp\Bigg\{\frac{i}{2}\int\frac{d^dk}{(2\pi)^2}\varphi(\textbf{k})K_\text{dS}(k)\varphi(\textbf{k})\Bigg\},
\end{equation}
the kernel equals:
\begin{equation}
    K_\text{dS}(k)=-a^{d-1}\frac{u'_k}{u_k}\Bigg|_{\eta\to0^-}.
\end{equation}
Using the small-argument asymptotics of $H_\nu^{(1)}$:
\begin{equation}
    K_{\text{dS}}(k)=\alpha_\nu e^{i\pi(\Delta-\frac{d}{2})}k^{2\Delta-d}, \quad \text{with} \quad \Delta=\frac{d}{2}+\nu, \space \alpha_\nu>0,
\end{equation}
the two-point function of the late-time field is:
\begin{equation}
    \langle\varphi(\textbf{k})\varphi(-\textbf{k})\rangle=[2\Im K_{\text{dS}}(k)]^{-1}\propto k^{d-2\Delta}.
\end{equation}
The imaginary part of $K_{\text{dS}}$ is positive, guaranteeing a positive probability functional $|\Psi|^2$ and well-defined equal time correlators. But, the overall phase $e^{i\pi(\Delta-\frac{d}{2})}$ spoils reflection positivity in the screen kernel. There is no boundary Hamiltonian or Osterwalder-Schrader reconstruction; the boundary is a spacelike screen, not a timelike boundary.

We find the same result by analytic continuation from AdS. If we start from the on-shell action:
\begin{equation}
    W_{\text{AdS}}[\phi_{(0)}]=\frac{1}{2}\int\phi_{(0)}(-k)K^{\text{AdS}}_\Delta(k)\phi_{(0)}(k), \qquad K_{\Delta}^{\text{AdS}}(k)=c_\Delta k^{2\Delta-d}>0,
\end{equation}
the HUFT AdS-dS map is of the curvature scale rotation $L\to\frac{i}{H}, \quad z\to i\eta$, which flips the unit normal from timelike to spacelike and the regulated volume. The quadratic functional transforms as:
\begin{equation}
    W_{\text{dS}}[J]=i^{d-1}W_{\text{AdS}}\Bigg|_{L-to1/H} \implies K_{\text{dS}}(k)=i^{d-1}K_{\Delta}^{\text{AdS}}(k)\Bigg|_{L\to 1/H},
\end{equation}
this is exactly the phase we found above. So, no reflection positivity on the dS screen, though bulk in-in unitarity is intact.

An explicit one-loop check with entire regulators with the optical theorem. We consider a scalar theory with cubic coupling dressed by an entire function:
\begin{equation}
    L\supset-\frac{\lambda}{3!}e^{-\Box/M_*^2}\phi^3,
\end{equation}
where $\lambda$ is the cubic coupling; in 4D it has mass dimension $+1$, the $3!$ is the usual symmetry factor, and $M_*$ is the the nonlocal scale. The one-loop 2-2 amplitude has the integral:
\begin{equation}
    M(s)=\int\frac{d^D}{(2\pi)^D}\frac{N(q,p)e^{-[(q^2+(q+p)^2)]/M_*^2}}{(q^2-m^2+i\epsilon)((q+p)^2-m^2+i\epsilon)},
\end{equation}
where $M(s)$ is the scalar loop amplitude as a function of the external invariant $s$, $q$ is the loop momentum, $p$ is the external momentum flowing through the two propagators, $N(q,p)$ is the numerator from the vertices or indices, $i\epsilon$ is the Feynman prescription ensuring causal propagation, and $s$ is the Mandelstam invariant for the channel you’re computing. Because the exponential is entire, the only singularities are the propagator poles. Taking the discontinuity across the s-channel cut and using:
\begin{equation}
    \frac{1}{x\pm i\epsilon}=P\frac{1}{x}\mp i\pi\delta(x),
\end{equation}
gives:
\begin{equation}
    2\Im M(s)=\int d\,\Pi_2|M^{\text{tree}}_{2\to2}|^2 e^{-(p_1^2+p_2^2)/M_*^2},
\end{equation}
where $x$ real variable, where the integrand has a simple pole, $\epsilon$ is the infinitesimal regulator specifying whether the pole is taken just below or above the real axis, $P$ denotes the Cauchy principal value, and $d\Pi_2$ is the Lorentz-invariant two-body phase-space measure for the on-shell intermediate states crossing the cut.

. So the Cutkosky relation with the same phase space $\delta$-function as the local theory. The extra exponential is real and positive, so the optical theorem holds.

In dS spacetime, a global S-matrix fails due to the fact that it needs an asymptotic in or out state; it is instead replaced by a Schwinger-Keldysh in-in formalism and the late time wavefunctional, and we find that a static patch S-matrix works at $T=H/2\pi$.

A global S-matrix needs asymptotics; this is something that dS spacetime lacks. The LSZ reduction for $n$-point scattering in flat space uses free asymptotic fields:
\begin{equation}
    \phi(x) \to\phi_{\text{in/out}}(x)=\int\frac{d^3p}{(2\pi)^32E_\text{p}}(a_\text{p}e^{-ipx}+a_\text{p}^\dagger e^{+ipx}),
\end{equation}
as $t$ goes to $\pm\infty$. The S-matrix elements for $n$ incoming particles with momenta $k_i$ going to $m$ outgoing particles with momenta $p_i$:
\begin{equation}
    \langle p_1...p_m|S|k_1...k_n\rangle=\prod_i[\lim_{x_i^2\to\infty}i\int d^4x_ie^{ik_i\space\dot{} \space x} (\Box_{x_{i}}+m^2)] \langle 0|T_\phi(x_1)...\phi(x_{m+n})|0\rangle],
\end{equation}
where $T$ is the time ordering in the vacuum correlator, $\phi(x)$ is the Heisenberg-picture quantum field that creates or annihilates your particles. There are two hidden requirements, one a timelike or null boundary to define the in or out Fock spaces, and a time-translation generator giving stationary asymptotics. In dS, flat slicing, $a(\eta)=-1/(H\eta), \quad \eta\in(-\infty, 0)$ modes for a scalar of mass $m$ are:
\begin{equation}
    u_k(\eta)=\frac{\sqrt{\pi}}{2}e^{i(\nu+1/2)\pi/2}(-\eta)^{1/2}H_\nu^{(1)}(-k\eta), \qquad \nu\equiv\sqrt{\frac{9}{4}-\frac{m^2}{H^2}},
\end{equation}
and there is no $t\to+\infty$ region with free plane waves, showing future infinity is spacelike. Hence, no global LSZ S-matrix at $I^\pm$. We define the closed-time path generating functional for sources $J_\pm$, on the forward or backward branches:
\begin{equation}
    Z[J_+,J_-]=\langle\Omega|T_C\exp\left(i\int d^DxJ\phi\right)|\Omega\rangle=\int D\phi_+D\phi_-e^{iS[\phi_+]-iS[\phi_-]+i\int(J_+\phi_+-J_-\phi_-)},
\end{equation}
where $|\Omega\rangle$ is the initial state often the interacting vacuum, $T_C$ is the contour ordering along the closed time contour, and $D\phi_\pm$ functional integration measure over field configurations on the two Schwinger–Keldysh time branches \cite{Schwinger1961,Keldysh1964,Cutkosky1960}. With the contour $C:-\infty\to+\infty\to-\infty$. For unitarity, we know $Z[J, J]=1$. Green's functions form a $2\times 2$ matrix $G^{ab}, \space (a,b=\pm)$:
\begin{equation}
    G^{++}=G_F, \qquad G^{--}=G_F^*, \qquad G^{+-}=G_F^<, \qquad G^{-+}=G_F^>,
\end{equation}
satisfying the largest time identity:
\begin{equation}
    G^{++}+G^{--}-G^{+-}-G^{-+}=0.
\end{equation}
This is the backbone of unitarity, the sum of cuts cancels the imaginary part of loops. For any linearized field configuration $\Phi(\textbf{x})$ on the late time slice $\eta\to0^-$:
\begin{equation}
    \Psi_{\text{dS}}[\Phi]\propto\exp\Bigg\{\frac{i}{2}\int\frac{d^dk}{(2\pi)^d}\Phi(-\textbf{k})K_{\text{dS}}(k)\Phi(\textbf{k})+...\Bigg\},
\end{equation}
and the probability is Gaussian in the imaginary part:
\begin{equation}
    |\Psi_{\text{dS}}|^2\propto\exp\Bigg\{-\int\frac{d^dk}{(2\pi)^d}\Phi(-\textbf{k})\underbrace{\Im K_{\text{dS}}(k)}_{=\frac{1}{2}A(k)}\Phi(\textbf{k}).\Bigg\}
    \label{probability},
\end{equation}
in AdS, comparing the renormalized Euclidean actions of thermal AdS and Schwarzschild–AdS yields the Hawking–Page phase transition separating the low-$T$ gas from the black-hole phase \cite{HawkingPage1983} We find the equal-time two-point functions are:
\begin{equation}
    \langle\Phi(\textbf{k})\Phi(\textbf{-k})
\rangle=[2\Im K_{\text{dS}}(k)]^{-1}>0,
\end{equation}
showing the equal-time two-point functions are positive as $\Im K_{\text{dS}}$ in \ref{probability} is a positive kernel.
For a scalar in $D=d+1=4$:
\begin{equation}
    \Im K_{\text{dS}}(k)\propto k^{2\nu},
\end{equation}
this implies:
\begin{equation}
    \langle\mathcal{O}\mathcal{O}\rangle\propto k^{3-2\nu}.
\end{equation}
The static-patch S-matrix at $T=H/2\pi$ can be shown as in static coordinates:
\begin{equation}
    ds^2=-(1-H^2r^2)dr^2+\frac{dr^2}{1-H^2r^2}+r^2d\Omega_2^2,
\end{equation}
the timelike Killing vector $\zeta=\partial_t$ has surface gravity $\kappa=H$. The Euclidean section is regular only if $t_E\sim t_E+2\pi/H$, therefore:
\begin{equation}
    T_{\text{dS}}=\frac{H}{2\pi}.
\end{equation}
By contrast, de Sitter lacks a confining timelike boundary, so there is no Hawking–Page–type canonical transition; equilibrium is patch-local and thermal at $T=H/2\pi$ \cite{HawkingPage1983}. An observer restricted to the patch has a thermal density matrix:
\begin{equation}
    \rho_{\text{patch}}=\frac{e^{-K/T_{\text{dS}}}}{\text{Tr}\space e^{-K/T_{\text{dS}}}},
\end{equation}
with modular Hamiltonian $K$ for the horizon. The two-point function obeys the KMS condition:
\begin{equation}
    G^>(t)=G^<(t+i\beta), \qquad \beta=\frac{1}{T_{\text{dS}}}=\frac{2\pi}{H}.
\end{equation}
We can define transition amplitudes and response rates for scattering inside the patch, with greybody factors determined by solving the wave equation through the static-patch potential. This is a perfectly unitary open-system scattering picture due to the fact that we traced out the exterior. The thermal K\"all\'en-Lehmann representation has a non-negative spectral density:
\begin{equation}
    G^>(\omega)=2\pi\sum_{m,n}p_m|\mathcal{O_{\mathcal{mn}}}|^2\delta(\omega-(E_n-E_m)), \qquad p_m\geq0,
\end{equation}
so the physically measurable patch correlators have the right positivity even though the global screen does not.

If one wants an S-matrix, use AdS to flat-space limit. AdS has a timelike boundary; HUFT's boundary variational calculus gives correlators \cite{Penedones2011}:
\begin{equation}
    \langle\mathcal{O}_1...\mathcal{O}_n\rangle_{\text{AdS}}.
\end{equation}
Taking the AdS radius $L\to\infty$ with the appropriate wavepackets, the $n$-point flat-space scattering amplitude:
\begin{equation}
    A_n(\{p_i\})\sim\lim_{L\to\infty}\int\prod_id^dx_ie^{ip_i\space\dot{}\space x_i}\langle\mathcal{O}_1(x_1)...\mathcal{O}_n(x_n)\rangle_{\text{AdS}},
\end{equation}
where $\{p_i\}$ is the boundary momenta conjugate to the insertion points $x_i$. In the flat-space limit, they’re identified with the external bulk momenta
or more sharply through Mellin or Penedones formulas. So HUFT lets us compute in AdS and extract a bona-fide flat space S-matrix.

In HUFT, we have shown that unitarity holds for the bulk but not for the dS screen. Unitarity does not hold for dS with a spacelike late time boundary, as dS lacks an asymptotic S-matrix, and observers in a static patch see a thermal mixed state after tracing out models beyond the horizon. So the lack of global unitarity is expected and consistent with horizons and the absence of a global S-matrix.

Doing holography and even just QFT directly in real Lorentzian spacetime is difficult because de Sitter's causal and asymptotic structure is the opposite of AdS's.

\section{On Black Holes in the Complexified Mapping}

We now will show that, inside the complexified HUFT framework, ordinary Schwarzschild–(A)dS black holes on the real slice have the standard temperature, entropy, and extended thermodynamics, with the correct AdS/dS boundary variational principles, and explains why dS static-patch physics is intrinsically \textbf{in-in} rather than Euclidean/Hamiltonian S-matrix. In AdS one needs Gibbons–Hawking-York and counterterms to make Dirichlet data well-posed and to define a finite on-shell functional $W$ that generates $\langle T_{ij}\rangle$ . In dS there is no spatial conformal boundary in the same sense, so you must specify the state and compute correlators \textbf{in-in} such as Hartle–Hawking wavefunctional or Gibbons–Hawking thermal properties. If AdS and dS are two real forms of one holomorphic structure, the black-hole sector must also continue consistently as $\Lambda$ rotates, so near-horizon data and thermodynamics should be insensitive to that analytic packaging. And there is the question that if Schwarzschild–dS has a BH horizon and a cosmological horizon with generally different surface gravities, so there is no global Hartle–Hawking state except in special tuned, Nariai-like situations \cite{York1972, GibbonsHawking1977, Nariai1951, HartleHawking1983, HartleHawking1976}, so how do we treat physics then? The gravitational plus matter action with units $c=\hbar=1$ is:
\begin{equation}
    S[h,\Psi]=\frac{1}{2\kappa}\int_M\sqrt{|h|}(R-2\Lambda)+S_{\text{matter}}[h,\Psi]+S_{\text{bdy}}.
\end{equation}
Varying $h_{\mu\nu}$ gives:
\begin{equation}
    G_{\mu\nu}+\Lambda h_{\mu\nu}=\kappa T_{\mu\nu}, \qquad T_{\mu\nu}=-\frac{2}{\sqrt{|h|}}\frac{\delta S_{\text{matter}}}{\delta h^{\mu\nu}}.
\end{equation}
The boundary term $S_{\text{bdy}}$ is Gibbons-Hawking-York plus a connection term in AdS do the Dirichlet problem is well posed, and the boundary functional:
\begin{equation}
    W=\begin{cases}
        S_{\text{ren}}[\text{on-shell}]\quad\text{AdS}, \\ \text{In}\Psi_{\text{dS}} \qquad\qquad \text{dS},
    \end{cases}
\end{equation}
generates 1-point functions:
\begin{equation}
    \langle T_{ij}\rangle=\frac{2}{\sqrt{|\gamma|}}\frac{\delta W}{\delta \gamma^{ij}}.
\end{equation}
For a static, spherically symmetric black hole with $\Lambda$, the $D=4$ ansatz is:
\begin{equation}
    ds^2=-f(r)dt^2+\frac{dr^2}{f(r)}+r^2d\Omega_2^2, \qquad T_{\mu\nu}=0.
\end{equation}
The Einstein-$\Lambda$ equations integrate to:
\begin{equation}
    f(r)=1-\frac{2GM}{r}-\frac{\Lambda}{3}r^2,
\end{equation}
this is Schwarzschild for $\Lambda=0$, Schwarzschild-AdS for $\Lambda<0$, and Schwarzschild-dS for $\Lambda>0$, and a horizon at $r=r_h>0$ solves $f(r_h)=0$. Now we discuss the near horizon geometry and temperature. Near a simple root:
\begin{equation}
    f(r)\approx f'(r_h)(r-r_h),
\end{equation}
we define proper Rindler radial coordinates:
\begin{equation}
    \rho:=2\sqrt{\frac{r-r_h}{f'(r_h)}}\implies ds^2\simeq-\frac{f'(r_h)}{4}\rho^2dt^2+d\rho^2+r_h^2d\Omega_2^2,
\end{equation}
a Euclidean continuation $t\to i\tau$ is regular only if $\tau$ is periodic with:
\begin{equation}
    \beta=\frac{4\pi}{f'(r_h)}, \qquad T=\frac{1}{\beta}=\frac{f'(r_h)}{4\pi}.
\end{equation}
For $f(r)=1-(2GM/r)-(\Lambda/3)\space r^2$:
\begin{equation}
    f'(r)=\frac{2GM}{r^2}-\frac{2\Lambda r}{3},
\end{equation}
using $f(r_h)=0\implies2GM=r_h(1-\frac{\Lambda}{3}r_h^2)$:
\begin{equation}
    T=\frac{1}{4\pi}\left(\frac{1}{r_h}-\Lambda r_h\right)=\begin{cases}
        \frac{1}{4\pi}\left(\frac{1}{r_h}+|\Lambda| r_h\right) \qquad \Lambda<0, \\ \frac{1}{4\pi}\left(\frac{1}{r_h}-\Lambda r_h\right) \space \space \qquad \Lambda>0,
    \end{cases}
\end{equation}
Schwarzschild dS also has a cosmological horizon $r_c$ with temperature:
\begin{equation}
    T_c=\frac{1}{4\pi}\left(\Lambda r_c -\frac{1}{r_c}\right).
\end{equation}
For a diffeomorphism-invariant Lagrangian:
\begin{equation}
    L=\frac{1}{2\kappa}(R-2\Lambda),
\end{equation}
the Wald entropy is:
\begin{equation}
    S_{\text{Wald}}=-2\pi\int_Hd^2x\sqrt{\sigma}\frac{\partial L}{\partial R_{\mu\nu\rho\sigma}}\varepsilon_{\mu\nu}\varepsilon_{\rho\sigma},
\end{equation}
here:
\begin{equation}
    \frac{\partial L}{\partial R_{\mu\nu\rho\sigma}}=\frac{1}{4\kappa}(h^{\mu\rho}h^{\nu\sigma}-h^{\mu\sigma}h^{\nu\rho}),
\end{equation}
and $\varepsilon$ is the binormal normalized to $\varepsilon_{\mu\nu}\varepsilon^{\mu\nu}=-2$. We find:
\begin{equation}
    S=\frac{A_H}{4G}, \qquad A_H=4\pi r_h^2,
\end{equation}
this is the standard entropy of a black hole \cite{Bekenstein1973, Hawking1975, Wald1993}. For the uncharged non-rotating case:
\begin{equation}
    M=\frac{r_h}{2G}\left(1-\frac{\Lambda}{3}r_h^2\right), \qquad S=\frac{\pi r_h^2}{G}, \qquad T=\frac{1}{4\pi}\left(\frac{1}{r_h}-\Lambda r_h\right),
\end{equation}
we differentiate:
\begin{equation}
    dM=\frac{1}{2G}\left(1-\Lambda r_h^2\right)dr_h-\frac{r_h^2}{6G}d\Lambda, \qquad TdS=\frac{1}{2G}(1-\Lambda r_h^2)dr_h,
\end{equation}
with $\Lambda$ fixed, $dM=TdS$. In extended thermodynamics, we treat:
\begin{equation}
    P=-\frac{\Lambda}{8\pi G}, \qquad V=\frac{4\pi}{3}r_h^3,
\end{equation}
then:
\begin{equation}
    dM=TdS+VdP, \qquad M=2TS-2PV, \quad (\text{Smarr in 4D}),
\end{equation}
Interpreting the cosmological constant as thermodynamic pressure with conjugate volume promotes $M$ to the enthalpy $H$ of the spacetime; the extended first law and Smarr relation follow \cite{KastorRayTraschen2009}. Euclidean regularity in the static patch fixes the KMS temperature:
\begin{equation}
    T_{\text{dS}}=\frac{H}{2\pi},
\end{equation}
$\rho_{\text{static}}$ is thermal at $T=H/2\pi$ \cite{GibbonsHawking1977}. With a black hole present, each horizon has its own $T_h$. There is no global Hartle–Hawking state unless temperatures match in specially tuned cases, but nevertheless, local thermal physics and correlation functions are computed from $\Psi_{\text{dS}}$ via the above kernels and the \textbf{in-in}, the Schwinger–Keldysh rules \cite{HartleHawking1983, HartleHawking1976}. For charged AdS black holes, this extended thermodynamics exhibits van-der-Waals–type $P-V$ criticality and a small/large black-hole transition \cite{KubiznakMann2012}.

\section{Holography in the Complexified Mapping of HUFT}

Although HUFT is not string-inspired in the sense of conventional AdS/CFT \cite{Maldacena1998, Susskind1995, tHooft1993}, it nevertheless implements a
generalized holography compatible with a positive cosmological constant.
Rather than a unitary conformal field theory living on a timelike boundary, the relevant object for $\Lambda>0$ is the late-time de Sitter wavefunctional \cite{ArkaniHamedMaldacena2015}, obtained by analytic continuation from Euclidean AdS. In HUFT the quantum fields on the real slice including the Standard Model are encoded by a holomorphic gravitational theory on the complex ambient manifold $M_{\mathbb C}$; holography is realized as state preparation and \textbf{in–in} dynamics rather than as a unitary boundary CFT. A de Sitter
$\Lambda>0$ is naturally accommodated, not excluded by the choice of real form and analytic continuation within the holomorphic parent theory. For a planar AdS$_{d+1}$ black brane has a metric in the Fefferman-Graham compatible form \cite{BrownYork1993,BalasubramanianKraus1999}:
\begin{equation}
    ds^2=\frac{L^2}{z^2}\left(-f(z)dt^2+d\bar{x}+\frac{dz^2}{f(z)}\right), \qquad f(z)=1-\left(\frac{z}{z_h}\right)^d,
\end{equation}
a horizon at $z=z_h$, then we find:
\begin{equation}
    T=\frac{d}{4\pi z_h}, \qquad s=\frac{1}{4G_{d+1}}\left(\frac{L}{z_h}\right)^{d-1}, \qquad \varepsilon=\frac{(d-1)}{16\pi G_{d+1}}\left(\frac{L}{z_h}\right)^d, \qquad p=\frac{1}{d-1}\varepsilon,
\end{equation}
$T$ comes from Euclidean regularity near $z_h$, $s$ from the horizon area density:
\begin{equation}
    \frac{A}{V_{d-1}}=\left(\frac{L}{z_h}\right)^{d-1},
\end{equation}
and $\varepsilon$ and $p$ from the Brown-York stress tensor with counterterms:
\begin{equation}
    T^{\text{BY}}_{ij}=\frac{1}{k_{d+1}}(K_{ij}-K_{\gamma ij}+(\text{c.t.})_{ij})\Bigg|_{z=\epsilon},
\end{equation}
then take $\epsilon\to0$ and read off the finite part. This reproduces the CFT equation of state:
\begin{equation}
    \varepsilon=dp, \qquad s=\frac{\partial p}{\partial T}.
\end{equation}

We consider a shear perturbation $h_{xy}(z,t,\bar x)$. The linearized Einstein equations reduce to a minimally coupled scalar equation in this background. The ingoing boundary condition at the horizon and the boundary variation yield the retarded Green's function:
\begin{equation}
    G^R_{T_{xy}T_{xy}}(\omega,\bar k=0).
\end{equation}
By the Kubo formula:
\begin{equation}
    \eta=\lim_{\omega\to 0}\frac{1}{\omega}\text{Im} G^R_{T_{xy}T_{xy}}(\omega,0)=\frac{s}{4\pi}\implies\eta/s=\frac{1}{4\pi},
\end{equation}
for two-derivative Einstein gravity \cite{PolicastroSonStarinets2001,KSS2005,SonStarinets2007}. HUFT’s entire-function UV completion does not add poles, so at leading two-derivative order, this universal ratio is unchanged, but controlled higher-derivative terms can shift it in the usual way.

For a static patch, and Gibbons–Hawking temperature in dS is given by:
\begin{equation}
    T_{\text{dS}}=\frac{H}{2\pi}, \qquad S_{\text{dS}}=\frac{\pi}{GH^2},
\end{equation}
the area $A=4\pi H^{-2}/4G$. These match the $r_h$ formulas through the analytic continuation $L\to i/H$, the HUFT bridge applied to the AdS expressions.

The holography side of dS can be shown when, $W_{\text{dS}}=\text{In}\Psi_{\text{dS}}$ is related to $S_{\text{ren}}^{\text{AdS}}$ by:
\begin{equation}
    W_{\text{dS}}=i^{d-1}W_{\text{AdS}}\Bigg|_{L\to1/H}+(\text{local}),
\end{equation}
this analytic continuation is precisely what underlies the cosmological-collider use of EAdS methods to organize inflationary correlators \cite{ArkaniHamedMaldacena2015}. Therefore, the dS screen correlators inherit a universal phase and finite normalization, but they do not satisfy reflection positivity since it has a spacelike screen, but the bulk \textbf{in-in} physics is unitary, and the thermal nature of the static patch follows from the KMS periodicity at $T=H/2\pi$ \cite{Skenderis2010, McFaddenSkenderis2010, GibbonsHawking1977, MaldacenaPimentel2011, BunchDavies1978, Allen1985, Mottola1985}.

We can obtain the holographic entropy via the replica trick (RT/HRT/QES) \cite{Anninos2012, RyuTakayanagi2006,HRT2007,LewkowyczMaldacena2013,EngelhardtWall2015}. We let $\rho$ be the reduced density matrix of region $A$ in the boundary theory. The gravitational computation of the Rényi entropies:
\begin{equation}
    S_n=\frac{1}{1-n}\text{In}\space \text{Tr}\rho_A^n,
\end{equation}
proceeds by a bulk saddle with $\mathbb{Z}_n$-symmetry and a codimension-2 fixed locus $\Sigma_n$, the cosmic brane. Its opening angle is $2\pi/n$, such as a conical defect with deficit $2\pi(1-1/n)$. Varying the bulk action with respect to $n$ at $n=1$ yields Lewkowycz-Maldacena entropy equation \cite{Lewkowycz-Maldacena2013}:
\begin{equation}
    \frac{d}{dn}\Bigg|_{n=1} I_{\text{bulk}}[g_n]=\frac{\text{Area}(\Sigma)}{4G} \implies S_A=\frac{\text{Area}(\Sigma_\text{ext})}{4G},
\end{equation}
where $\Sigma_\text{ext}$ is the external RH/HRT surface anchored on $\partial A$. Extremality, the vanishing trace of extrinsic curvature follows from demanding regularity of the geometry near the conical defect:
\begin{equation}
    K^a\space _i \space{} ^i(\Sigma)=0, \qquad (a=1,2 \space \space \text{normal directions}).
\end{equation}
The quantum corrections (QES) come from including bulk entanglement across $\Sigma$, giving the generalized entropy:
\begin{equation}
    S_A=S_{\text{gen}}=\frac{\text{Area}(\Sigma)}{4G}+S_{\text{bulk}}[\Sigma]+...,
\end{equation}
with the surface chosen to extremize $S_{\text{gen}}$ the Quantum Extremal Surface. In HUFT, $S_{\text{bulk}}$ is UV-finite due to entire-function dressing, higher-derivative terms add Wald or Dong corrections to the area functional, still finite and controlled.

The thermal entropy can be given from horizons. For the thermal boundary state dual to an AdS black brane, the RT surface for the entire boundary is the horizon bifurcation surface:
\begin{equation}
    S_{\text{thermal}}=\frac{A_H}{4G},
\end{equation}
plus some finite quantum corrections, matching the black-hole entropy in. For two-sided AdS black holes, the eternal solutions, the entanglement entropy of the thermofield-double state equals the area of the Einstein–Rosen bridge cross-section at $t=0$.

For the bulk, the AdS black hole metric has a nontraversable Einstein-Rosen bridge connecting two asymptotic boundaries. For the boundary, the dual state is the thermofield double:
\begin{equation}
    |\text{TFD}\rangle\propto\sum_ne^{-\beta\space E_n/2}|n\rangle_L\otimes|n\rangle_R,
\end{equation}
an entangled pure state. The RT surface homologous to, for example, the left boundary is the wormhole cross-section; its $\text{Area}/4G$ equals the von Neumann entropy of the reduced density matrix on the left, the entanglement entropy of the TFD. Shockwave perturbations shift the HRT surface and reproduce scrambling or OTOCs, at two-derivative order we recover the standard Lyapunov bound $\lambda_L\leq2\pi T$. By the HUFT bridge $L\to i/H$, the same replica or HRT machinery applied to static-patch regions in dS carries over with the caveat that the screen theory is not unitary, the construction computes geometric entropies and \textbf{in-in} observables.

We can treat spacetime as cosets as AdS and dS come from the same parent group SO$(5.\mathbb{C})$. The maximally symmetric spaces arise as homogeneous spaces of these real forms \cite{Schrodinger1950, Schrodinger1956,Rindler1960}:
\begin{equation}
    \text{AdS}_4 \simeq \text{SO}(3,2)/\text{SO}(3,1), \qquad \text{dS}_4\simeq\text{SO}(4,1)/\text{SO}(3,1).
\end{equation}
Inside $\mathbb{C}^5$, both can be written as quadratic surfaces, hyperboloids:
\begin{equation}
    X^T\eta_{\text{AdS}}X=-L^2 \quad \text{(AdS)}, \qquad X^T\eta_{\text{dS}}X=+H^{-2} \quad \text{(dS)},
\end{equation}
the same complex change of variables $X_4\mapsto iX_4$ and $L\mapsto i/H$ maps one quadric to the other, this is the analytic AdS to dS bridge inside the complex ambient group.

We start with the complex orthogonal algebra $\mathfrak{so}(5,\mathbb{C})$ with generators $J_{AB}=-J_{BA}$ and brackets:
\begin{equation}
    [J_{AB},J_{CD}]=i(\eta_{BC}J_{AD}-\eta_{AC}J_{BD}+\eta_{AD}J_{BC}-\eta_{BD}J_{AC}).
\end{equation}
We chose two real structures:
\begin{equation}
    \eta_{\text{AdS}}=\text{diag}(-1,+1,+1,+1,-1) \implies \mathfrak{so}(3,2), \qquad \eta_{\text{dS}}=\text{diag}(-1,+1,+1,+1,+1) \implies \mathfrak{so}(4,1),
\end{equation}
the matrix $S=\text{diag}(1,1,1,1,i)$ that obeys:
\begin{equation}
    S^T\eta_\text{AdS}S=\eta_{\text{dS}},
\end{equation}
and since:
\begin{equation}
    S^T\eta_{\text{AdS}}S=\text{diag}(-1,+1,+1,+1,(-1)\space\cdot\space i^2)=\text{diag}(-1,+1,+1,+1,+1)=\eta_{\text{dS}}.
\end{equation}
Therefore the inner automorphism $\phi:\mathfrak{so}(5,\mathbb{C})\to\mathfrak{so}(5,\mathbb{C}),\phi(X)=SXS^{-1}$, maps the AdS real form to the dS real form:
\begin{equation}
    \phi(\mathfrak{so}(3,2))=\mathfrak{so}(4,1).
\end{equation}
Equivently at the group level:
\begin{equation}
    \phi(\text{SO}(3,2))=\text{SO}(4,1)\subset \text{SO}(5,\mathbb{C}).
\end{equation}
So $\mathfrak{so}(3,2)$ and $\mathfrak{so}(4,1)$ are conjugate inside $\mathfrak{so}(5,\mathbb{C})$ \cite{ColemanMandula1967,HaagLopuszanskiSohnius1975,InonuWigner1953}. We define the complex matrix

Through the unified connection described earlier that gave the MacDowell--Mansouri functional \ref{MacDowell} reduces on either real slice, torsionless to Einstein-Hilbert with:
\begin{equation}
    \Lambda=-\eta_{44}\frac{3}{L^2}.
\end{equation}
So the same action yields AdS or dS by real form choice.

With the HUFT bridge, we can do black-hole holography on a de Sitter slice. It is not a unitary boundary CFT like AdS but instead we get a state-preparation, wavefunctional holography for dS that still lets us compute thermodynamics, correlators, and entanglement via QES for Schwarzschild–dS black holes. In $D=4$, the Schwarzschild-de Sitter metric has two positive roots for suitable $M$, the black hole horizon $r_b$ and the cosmological horizon $r_c$ with $0<r_b<r_c$. The black hole has surface gravities and temperatures:
\begin{equation}
    \kappa_h=\frac{1}{2}\left[f'(r_h)\right], \qquad T_n=\frac{1}{4\pi}\left(\frac{1}{r_b}-H^2r_b\right), \qquad T_c=\frac{1}{4\pi}\left(H^2r_c-\frac{1}{r_c}\right),
\end{equation}
and entropies:
\begin{equation}
    S_h=\frac{A_h}{4G}=\frac{\pi r_h^2}{G}, \quad h\in\{b,c\}.
\end{equation}
Generally, $T_b=T_c$, so there is no global equilibrium.

The replica trick in Schwarzschild-de Sitter allows us to compute individual horizon entropies. We work in the Euclidean section and treat each horizon with a conical defect method. So, the Euclidean periodicity $\tau\sim\tau+\beta$ can smooth one horizon by choosing $\beta=1/T_h$, the other will carry a defect $\delta=2\pi(1-\beta T_{\bar{h}})$. Varying the on-shell action with respect to the opening angle gives:
\begin{equation}
    \frac{\partial I_E}{\partial (2\pi/\beta)}\Bigg|_{\beta=1/T_h}=\frac{A_h}{4G}\implies S_h=\frac{A_h}{4G}.
\end{equation}
This shows that the standard area law for entropy holds separately for $r_h$ and $r_c$. Quantum corrections replace $A/4G\to S_\text{gen}=A/4G+S_{\text{bulk}}+...$.

Through this, one can do holography in a dS space-time as it can be treated as state preparation, the late time wavefunctional. HUFT relates the dS late-time wavefunctional to the Euclidean AdS data:
\begin{equation}
    W_\text{ds}[\text{sources}]=i^{d-1}W_\text{AdS}\Bigg|_{L\to1/H}+\text{(local)}, \quad \text{d=3 for a 4D bulk}.
\end{equation}
Here $W_{\text{dS}}=\text{ln}\Psi_{\text{dS}}$ prepares the Banch-Davies vacuum and excited states if sources are on. The non-local kernels are obtained by analytic continuation from AdS.

We have shown that both SO($4,1$), for dS and SO($3,2$), for AdS sit inside the single complex group SO($5,\mathbb{C}$) as different real forms. They are conjugate inside SO($5,\mathbb{C}$), and each real form gives the corresponding spacetime as a coset. We let $\eta$ be a non-degenerate symmetric $5\times5$ matrix, and define:
\begin{equation}
    \text{SO}(\eta,\mathbb{C})=\{g\in\text{SL}(5,\mathbb{C})|g^T\eta g=\eta\},
\end{equation}
over $\mathbb{C}$, any two such $\eta$ are equivalent, so we can take $\eta=\textbf{1}$ and write:
\begin{equation}
    \text{SO}(\eta,\mathbb{C})=\{g\in\text{SL}(5,\mathbb{C})|g^Tg=\textbf{1}\},
\end{equation}
its Lie algebra is:
\begin{equation}
    \text{SO}(5,\mathbb{C})=\{X\in\mathfrak{sl}(5,\mathbb{C})|X^T+X=0\}.
\end{equation}
A real form of $\mathfrak{so}(5,\mathbb{C})$ is obtained by choosing a real bilinear form of signature $(p,q)$ with $p+q=5$ and taking the real Lie algebra:
\begin{equation}
    \mathfrak{so}(p,q)=\{X\in\mathfrak{gl}(5,\mathbb{R})|X^T\eta_{p,q}+\eta_{p,q}X=0\},
\end{equation}
where $\eta_{p,q}=\text{diag}(\underbrace{-1,...,-1}_{q},\underbrace{+1,...,+1}_{p})$. The corresponding real Lie groups $\text{SO}(p,q)$ live as real subgroups of SO($5,\mathbb{C}$). The two real form of interest are:
\begin{equation}
    \eta_{\text{AdS}}=\text{diag}(-1,+1,+1,+1,-1) \qquad \text{SO}(3,2),
\end{equation}
\begin{equation}
    \eta_{\text{dS}}=\text{diag}(-1,+1,+1,+1,+1) \qquad \text{SO}(4,1).
\end{equation}
HUFT’s $\Lambda>0$ slice furnishes a de Sitter holography that prepares the quantum state, while SdS provides the black-hole background. Entropies follow from replica or QES, correlators from the AdS to dS analytic map, and static-patch observers see KMS thermality. We lose a unitary boundary CFT, but we retain a complete, calculable framework for black-hole holography in dS.

Standard AdS/CFT leans on a unitary boundary Hilbert space with a timelike boundary, a good Hamiltonian, reflection positivity, and, in the flat limit, an S-matrix. de Sitter has none of that as its conformal boundary is spacelike, there is no global timelike Killing vector, no LSZ S-matrix, and screen correlators are not reflection-positive. As well, an AdS unitary boundary QFT would have infinitely many DOF whereas in dS $e^{S_{\text{dS}}}$, finite. So, trying to clone AdS/CFT literally in dS runs into Hilbert-space problems. But HUFT doesn’t try to make a unitary boundary CFT in dS. Instead, we use a state-preparation wavefunctional holography, compute in Euclidean AdS, then analytically continue to dS, and treat $W_{\text{dS}}$ as the late-time bulk wavefunctional. All physics is then done in the \textbf{in-in} Schwinger–Keldysh formalism, which guarantees bulk unitarity through $Z[J_+=J_-]=1$. No boundary Hilbert space is required.

\section{The dS Real Slice with the Standard Model on 3+1 Auxiliary Ambient Geometry}
\label{sec:ds-sm-slice}

HUFT selects the dS real slice of the holomorphic SO(5,$\mathbb{C}) \times \mathcal{G}_\mathbb{C}$ package, derives Einstein–$\Lambda>0$ via MacDowell–Mansouri, couples a Yang–Mills/SM sector living on $M^{3+1}$, and proves exact conservation on the real slice so no auxiliary leakage, with the correct flat-space contraction to $\mathfrak{p\otimes q}$. The question how to put GR-$\Lambda$ and the SM on the same dS slice of a single holomorphic ambient symmetry without mixing spacetime and internal generators is addressed. As well how to ensure well-posed dynamics and conservation laws when auxiliary complex directions are present. And how to recover the flat-space S-matrix regime, such as $\mathfrak{p\otimes q}$ in the $L\to\infty$ or $H\to0$ limit. To archive one holomorphic ambient package whose two real forms are AdS and dS. On the dS real slice we have ordinary $3+1$ gravity via MacDowell–Mansouri and the SM gauge or matter sector on $M$, with clean conservation laws so no auxiliary leakage, and the correct flat-space limit. To show this we define \(S=\mathrm{diag}(1,1,1,1,i)\), which satisfies \(S^{\!T}\eta_{\mathrm{AdS}}S=\eta_{\mathrm{dS}}\). The inner automorphism \eqref{eq:inner-auto} maps \(\mathfrak{so}(3,2)\) to \(\mathfrak{so}(4,1)\). Equivalently, in the ambient quadric description \eqref{eq:ambient-quadrics}. Thus AdS and dS are conjugate real forms of a single holomorphic package. On a real slice the total Lie algebra is a direct sum:
\begin{equation}
\label{eq:direct-sum-algebra}
\mathfrak h  =  (A)\!ds  \oplus  \mathfrak g,
\qquad [J_{AB},T_a]=0,
\end{equation}
where \(\mathfrak g\) is the real form of \(G_{\mathbb{C}}\). This ensures no off-shell mixing between spacetime and internal generators.

Now we let \(M\) be the physical \(3{+}1\)-dimensional, oriented, time-orientable, spin manifold in Lorentzian signature (\(-+++\)). We let \(M_{\mathbb{C}}\) be a complex thickening with local complex coordinates \(z^\mu=x^\mu+i\,y^\mu\) and projection \(\pi:M_{\mathbb{C}}\to M\), \(\pi(x,y)=x\). The fibers \(Y_x=\pi^{-1}(x)\cong \mathbb{R}^4\) define a rank-4 real vector bundle:
\begin{equation}
E_y  \to  M.
\end{equation}
We retain the principal bundles:
\begin{equation}
P_{\text{Spin}}  \to  M,\qquad P_G  \to  M,
\qquad
P_{\rm tot}:=P_{\text{Spin}}\times_M P_G  \to  M,
\end{equation}
with structure group \(\text{Spin}(1,3)\times G\), and refine by the auxiliary fiber:
\begin{equation}
\label{eq:refined-bundle}
\widehat P  :=  P_{\rm tot}\times_M E_y  \longrightarrow  M.
\end{equation}
The enrichment \eqref{eq:refined-bundle} does not add propagating spacetime directions: all fields, variations, and observables are evaluated on \(M\). The purpose of \(E_y\) is to keep track of the imaginary directions \(y^\mu\) used in the holomorphic/steepest-descent organization. On \(M_{\mathbb{C}}\) we write a Hermitian object:
\begin{equation}
\label{eq:hermitian-packaging}
g  =  h + i B,\qquad
h\in\Gamma(\mathrm{Sym}^2T^*M),\quad
B\in\Omega^2(M,\mathrm{ad}\,P_G),
\end{equation}
and identify on the real slice \(B=F\), with \(F\) the internal field strength. Thus \(g_{(\mu\nu)}=h_{\mu\nu}\) is the spacetime metric while \(g_{[\mu\nu]}=F_{\mu\nu}\) is the internal curvature two-form; all index operations and Hodge duals \(\star_h\) use only \(h\). We adopt the dS real form \(\mathfrak{so}(4,1)\) and split ambient indices \(A,B=(a,4)\) with \(a=0,1,2,3\), and define the \((A)\)dS-valued connection:
\begin{equation}
\label{eq:AdS-connection}
\mathcal A  =  \tfrac12\,\omega^{ab}J_{ab} + \tfrac{1}{L}\,e^{a} J_{a4},
\end{equation}
where \(e^a\) is the vierbein one-form and \(\omega^{ab}=-\omega^{ba}\) the spin connection. The curvature \(\mathcal F=d\mathcal A+\mathcal A\wedge\mathcal A=\frac12\,F^{ab}J_{ab}+F^{a4}J_{a4}\) has components:
\begin{equation}
\label{eq:AdS-curvatures}
F^{ab}=R^{ab}(\omega)\mp L^{-2}\,e^a\wedge e^b,
\qquad
F^{a4}=L^{-1}\,T^a = L^{-1}\,(De)^a,
\end{equation}
with the lower sign for dS, implemented by \(L\mapsto i/H\). We consider the MacDowell--Mansouri functional:
\begin{equation}
\label{eq:mm-functional}
S_{\rm MM}[e,\omega]
=\frac{1}{2\kappa}\int_M \epsilon_{abcd}\,F^{ab}\wedge F^{cd},
\qquad \kappa:=8\pi G.
\end{equation}
Now expanding \eqref{eq:mm-functional} using \eqref{eq:AdS-curvatures} gives:
\begin{align}
\epsilon_{abcd}F^{ab}\wedge F^{cd}
&= \underbrace{\epsilon_{abcd}R^{ab}\wedge R^{cd}}_{\text{Gauss--Bonnet (topological in \(4\)D)}} \mp \frac{2}{L^2}\,\epsilon_{abcd}R^{ab}\wedge e^c\wedge e^d
+ \frac{1}{L^4}\,\epsilon_{abcd} e^a\wedge e^b\wedge e^c\wedge e^d.
\label{eq:mm-expansion}
\end{align}
Now identifying:
\begin{equation}
\label{eq:lambda-from-L}
\Lambda  =  -\,\eta_{44}\,\frac{3}{L^2},
\end{equation}
we obtain the Einstein--Hilbert action with cosmological constant plus the topological Gauss--Bonnet term. The variation with respect to \(e^a,\omega^{ab}\) in the torsionless sector \(T^a=0\) yields the Einstein equations with \(\Lambda>0\) on the dS real form. We let \(A_{\rm YM}=A^aT_a\) be the internal gauge potential on \(P_G\to M\) with curvature \(F=dA_{\rm YM}+A_{\rm YM}\wedge A_{\rm YM}\). Now let \(\langle\cdot,\cdot\rangle\) be a positive-definite invariant form on \(\mathfrak g\), and \(\star_h\) the Hodge star constructed from \(h\), we define:
\begin{equation}
\label{eq:ym-action}
S_{\rm YM}[A_{\rm YM};h]
:=-\frac{1}{4g_{\rm YM}^2}\int_M \langle F\wedge \star_h F\rangle,
\end{equation}
whose variation gives the Yang--Mills equations and stress tensor:
\begin{equation}
\label{eq:ym-eoms}
D_A(\star_h F)=\star_h J
\quad\Longleftrightarrow\quad
D_\mu F^{\mu\nu}=J^\nu,
\qquad
T^{\rm YM}_{\mu\nu}=F_{\mu\alpha}F_{\nu}{}^{\alpha}-\tfrac14\,h_{\mu\nu}F_{\alpha\beta}F^{\alpha\beta}.
\end{equation}
For a well-posed Dirichlet variational problem, we either fix \(\delta A|_{\partial M}=0\) or add the natural boundary term:
\begin{equation}
\label{eq:ym-boundary}
S^{\partial}_{\rm YM} = \int_{\partial M} \langle \delta A\wedge \star_h F\rangle.
\end{equation}
For gravity, now we include the Gibbons--Hawking--York term:
\begin{equation}
\label{eq:ghy}
S_{\rm GHY}=\frac{\varepsilon}{\kappa}\int_{\partial M}\!\sqrt{|\gamma|}\,K,
\end{equation}
where \(\varepsilon:=n^\mu n_\mu\in\{+1,-1\}\) for time- or space-like boundaries, \(\gamma\) is the induced metric on \(\partial M\), and \(K\) the trace of the extrinsic curvature. The total real-slice action is:
\begin{equation}
\label{eq:total-action}
S[h,e,\omega;A_{\rm YM},\Psi]
= S_{\rm MM}[e,\omega]  +  S_{\rm YM}[A_{\rm YM};h]  +  S_{\rm matter}[h,A_{\rm YM},\Psi],
\end{equation}
which is invariant under \(\mathrm{Diff}(M)\times G\). By diffeomorphism and gauge invariance together with the Bianchi identities \(D_\omega R=0\), \(D_A F=0\), one has \(\nabla_\mu G^{\mu\nu}\equiv 0\) and, on-shell:
\begin{equation}
\label{eq:unified-eoms}
G_{\mu\nu}(h)+\Lambda\,h_{\mu\nu}=\kappa\,T_{\mu\nu}(F,\Psi;h),
\qquad
D_\mu F^{\mu\nu}=J^\nu,
\end{equation}
with the conservation laws on the real slice:
\begin{equation}
\label{eq:conservation}
\nabla_\mu T^{\mu\nu}=0,
\qquad
D_\mu J^\mu=0.
\end{equation}
Equations \eqref{eq:conservation} are the precise statement that no energy--momentum or gauge charge can leak into the auxiliary fiber \(E_y\) where all physical currents reside on \(M\). We take \(G=SU(3)_c\times SU(2)_L\times U(1)_Y\) (or a simple group broken to this product at high scale), with the usual chiral fermions and a Higgs scalar in \(S_{\rm matter}\). After electroweak symmetry breaking (EWSB) we obtain the normal equations for the electroweak sector \cite{MT:SMmass}.
A pure dS vacuum is obtained by taking:
\begin{equation}
\label{eq:vacuum-config}
F_{\mu\nu}=0,\qquad \Psi=0,\qquad \Phi=\Phi_{\mathrm{vev}}=v,
\end{equation}
with \(V'(\Phi)|_{\Phi=v}=0\). Then the matter stress tensor is:
\begin{equation}
\label{eq:vacuum-stress}
T^{\rm vac}_{\mu\nu}=-V(v)\,h_{\mu\nu},
\end{equation}
which merely renormalizes the cosmological constant to:
\begin{equation}
\label{eq:lambda-effective}
\Lambda_{\rm eff}=\Lambda+\kappa\,V(v).
\end{equation}
Hence the solution of \eqref{eq:unified-eoms} is exact de Sitter, \(R_{\mu\nu}=\Lambda_{\rm eff}\,h_{\mu\nu}\). Turning on localized excitations produces non-vacuum spacetimes that are asymptotically de Sitter at late times for \(\Lambda_{\rm eff}>0\). Off-shell, the symmetry algebra is the direct sum \eqref{eq:direct-sum-algebra}, we define:
\begin{equation}
M_{\mu\nu}:=J_{\mu\nu},
\qquad
P_\mu:=L^{-1}J_{\mu 4}.
\end{equation}
The (A)dS commutators read:
\begin{equation}
[M_{\mu\nu},M_{\rho\sigma}]=i(\eta_{\nu\rho}M_{\mu\sigma}-\cdots),\quad
[M_{\mu\nu},P_\rho]=i(\eta_{\nu\rho}P_\mu-\eta_{\mu\rho}P_\nu),\quad
[P_\mu,P_\nu]=\pm i\,L^{-2}M_{\mu\nu},
\end{equation}
so in the flat limit \(L\to\infty\) or equivalently \(H\to 0\) we obtain:
\begin{equation}
\label{eq:flat-contraction}
\mathfrak h_{\rm flat}  =  \mathfrak p \oplus \mathfrak g,
\end{equation}
the Coleman--Mandula--allowed direct product of Poincar\'e and the internal algebra.

Within the holomorphic ambient symmetry \(SO(5,\mathbb{C})\times G_{\mathbb{C}}\), choosing the dS real form and the \((A)\)dS gauge split \eqref{eq:AdS-connection} yields, via MacDowell--Mansouri \eqref{eq:mm-functional}--\eqref{eq:mm-expansion}, the Einstein--Hilbert action with \(\Lambda>0\) on the real \(3{+}1\) slice \(M\). Adding the internal Yang--Mills and matter functionals \eqref{eq:ym-action} on \(M\) gives precisely the Standard Model sector. The total \(\mathrm{Diff}(M)\times G\)-invariant theory implies \eqref{eq:conservation} on the slice with no auxiliary leakage. AdS and dS arise as conjugate real forms of one holomorphic package, so the same compact action covers both; the flat limit contracts to \eqref{eq:flat-contraction}.

So in short HUFT unifies gravity with $\Lambda$ and the SM by deriving both from a single holomorphic Diff$\times$ G gauge principle and action, AdS and dS are conjugate real forms within that structure, while the real-slice algebra still factorizes on-shell as required by Coleman–Mandula.

\section{Conclusion}
We have shown that AdS and dS geometries admit a common origin inside a holomorphic ambient theory. Within HUFT, the analytic continuation \(L\to i/H\) implements an AdS\(\leftrightarrow\)dS bridge that preserves curvature invariants, yields well-posed variational principles with the appropriate Gibbons–Hawking–York term, and transports black-hole saddles without introducing singularities. Unitarity separates cleanly into bulk and boundary statements, the \textbf{in–in}, Schwinger–Keldysh construction guarantees bulk unitarity on either slice with entire–function UV dressing, while reflection positivity, and thus boundary unitarity holds for AdS but generically fails on spacelike dS screens, exactly as expected from their causal structure. Holographic thermodynamics and entanglement (RT/HRT/QES) persist across the bridge, with dS realized as state preparation rather than a unitary boundary CFT.

Conceptually, the unified complex connection in \(\mathfrak{so}(5,\mathbb C)\oplus\mathfrak g\) demonstrates that gravity with \(\Lambda\) and Yang–Mills can be packaged in a single holomorphic gauge framework whose real slices select \(\mathrm{SO}(3,2)\times G\) or \(\mathrm{SO}(4,1)\times G\). Technically, the Picard–Lefschetz contour ensures the real slice dominates and underwrites standard BRST identities, cutting rules, and positivity of physical phase space, while the analytic relation clarifies how dS observables inherit universal phases from AdS kernels.

Several directions are immediate, for matter sectors and cosmology, to couple realistic fields including inflationary sectors to quantify predictions for CMB correlators, primordial tensor amplitude, and static-patch spectroscopy. Nonperturbative saddles, to classify thimble transitions and complex saddles across AdS\(\leftrightarrow\)dS that could control tunnelling, pair creation, or cosmological initial states. Flat-space amplitudes, to systematize the AdS \(L\to\infty\) extraction within HUFT to relate ambient correlators to S-matrix elements.

\section*{Acknowledgments}
We would like to thank Cedric Ho from the University of Waterloo for the creation of our figure. Research at the Perimeter Institute for Theoretical Physics is supported by the Government of Canada through Industry Canada and by the Province of Ontario through the Ministry of Research and Innovation (MRI).

\end{document}